\documentstyle [aps,twocolumn,epsfig]{revtex}
\catcode`\@=11
\newcommand{\fcaption}[1]{\vspace{1ex}
        \refstepcounter{figure}
        \setbox\@tempboxa = \hbox{\footnotesize {\bf ig.~\thefigure.} #1}
        \ifdim \wd\@tempboxa > 8cm
           {\begin{center}
        \parbox{8cm}{\footnotesize\baselineskip=8pt {\bf Fig.~\thefigure.} #1}

            \end{center}}
        \else
             {\begin{center}
             {\footnotesize {\bf Fig.~\thefigure.} #1}
              \end{center}}
        \fi}
\newcommand{\be}{\begin{equation}}
\newcommand{\ee}{\end{equation}}
\newcommand{\bea}{\begin{eqnarray}}
\newcommand{\eea}{\end{eqnarray}}

\newcommand{\ri}{\mbox{i}}
\newcommand{\re}{\mbox{e}}

\begin{document}

\title{Metal-insulator transition in the one-dimensional SU(N) Hubbard model}
\vspace{2cm}

\author{Roland Assaraf$~^{1}$, Patrick Azaria$~^{2}$, Michel Caffarel$~^{1}$, 
and Philippe Lecheminant$~^{3}$}

\vspace{0.5cm}

\address{$^1$  CNRS-Laboratoire de Chimie Th\'eorique,
Universit{\'e} Pierre et Marie Curie, 4 Place Jussieu, 75252 Paris,
France \\
$^2$ CNRS-Laboratoire de Physique Th\'eorique des Liquides,
Universit\'e Pierre et Marie Curie, 4 Place Jussieu, 75252 Paris,
France \\
$^3$ Laboratoire de Physique Th\'eorique et Mod\'elisation,
Universit\'e de Cergy-Pontoise, Site de Saint Martin,
2 avenue Adolphe Chauvin, 95302
Cergy-Pontoise Cedex, France} 
\vspace{3cm}

\address{\rm (Received: )}
\address{\mbox{ }}
\address{\parbox{14cm}{\rm \mbox{ }\mbox{ }
We investigate the metal-insulator transition of the 
one-dimensional SU(N) Hubbard model for repulsive interaction. 
Using the bosonization approach a Mott transition in the charge sector 
at half-filling (k$_F$=$\pi/Na_0$) is conjectured for  $N > 2$. 
Expressions for the charge and spin velocities as well as for the 
Luttinger liquid parameters and some correlation functions are given. 
The theoretical predictions are compared with numerical results obtained with  
an improved zero-temperature quantum Monte Carlo approach. The method used 
is a generalized Green's function Monte Carlo scheme 
in which the stochastic time evolution is partially integrated out. 
Very accurate results for the gaps, velocities, and Luttinger liquid 
parameters as a function of the Coulomb interaction $U$ are given
for the  cases $N=3$ and $N=4$. Our results strongly support the existence of 
a Mott-Hubbard transition at a {\it non-zero} value of the Coulomb interaction. 
We find $U_c \sim 2.2$ for $N=3$  and $U_c \sim 2.8$ for $N=4$.}}
\address{\mbox{ }}
\address{\parbox{14cm}{\rm PACS No: 05.10.Ln, 71.10.Fd, 71.30.+h, 75.40.Gb}}
\maketitle
\section{Introduction}
Although the metal-insulator transition has certainly been one of the most studied 
phenomenon in condensed-matter physics, it is only 
in recent years that important progress has been achieved. 
This is mainly due to careful experimental and numerical studies but 
also to the improvement of the theoretical tools\cite{mott,gebhard,imada}. 
It has been proved extremely difficult to investigate the effect of strong correlations 
in dimensions greater than one, and it is only quite recently that, 
thanks to a new dynamical mean-field, our understanding has substantially 
progressed\cite{georges}. For one-dimensional systems, the situation is rather 
different: There exist powerful analytical and numerical approaches at our disposal.
Moreover, from the experimental point of view, the Mott-transition 
can be realized in organic conductors\cite{jerome} and quantum wires\cite{tarucha}. 
Therefore, one may expect to gain a lot of information on the physics
of the metal-insulator transition.

In one dimension, it has been 
recognized very rapidly that umklapp processes are at the heart of the
problem. In the Abelian bosonization formalism, one can draw a 
general and consistent picture of the Mott-transition. Indeed, 
the charge properties are expected to be described, in the absence 
of umklapp contributions, by a Luttinger liquid with only 
two independent parameters: The charge velocity $u_c$ 
and the charge exponent $K_c$ that controls the decay of correlation functions.
These quantities, which are non-universal, completely 
characterize the low-energy properties of a one-dimensional 
system\cite{haldane,voit}. Within this framework, the effect of 
umklapp processes are investigated in perturbation theory, 
and one can write down an effective theory that describes the 
Mott-transition as well as a  full description of the transport 
properties for any commensurate filling\cite{giamarchi,schulz}. 
The only parameter  that controls the transition is 
the (in general unknown) Luttinger charge exponent
$K_c$, and the transition is predicted 
to be universal of the Kosterlitz-Thouless (KT) type. 

Most of the theoretical work in $d=1$ focused on 
the properties of the standard SU(2) Hubbard model which,
is known to be a Mott insulator at half-filling from its exact 
solution \cite{lieb}.
Some extension of this model was considered by introducing long-range 
hopping or finite-range interaction (nearest neighbor
interaction for instance)\cite{gebhard}.
In the present work, we study a most natural 
generalization of the usual Hubbard model: Instead of considering 
fermions with a two-valued spin index (with a SU(2) symmetry) we 
generalize to the case of an arbitrary SU(N) spin index.
Apart from the theoretical interest it is important to emphasize that
these additional degrees of freedom are realized physically through 
orbital degeneracy as for example in  Mn oxides\cite{imada}.
In this paper, we shall study the phase diagram of the  
one-dimensional SU(N) Hubbard model for repulsive interaction and 
at half filling corresponding to  one ``electron'' per site.
The Hamiltonian on a finite chain with $L$ sites that we shall consider reads:
\be
{\cal H} = - t\; \sum_{i=1}^{L}\sum_{a=1}^N
\left(c_{ia}^{\dagger} c_{i+1a}
+ {\rm H.c.} \right) +
\frac{U}{2} \; \sum_{i=1}^L \left(\sum_{a=1}^N n_{ia}\right)^2
\label{hun}
\ee
where the fermion annihilation operator of spin index $a=1,..,N$
at site $i$ is denoted by $c_{ia}$ and satisfies the canonical
anticommutation relation:
\be
\{c_{ia}, c_{jb}^{\dagger} \} = \delta_{ab} \delta_{ij}.
\ee
The density of species $a$ at the i$^{th}$ site is defined
by: $n_{ia} = c_{ia}^{\dagger}c_{ia}$.
In the following, we shall consider that the nearest-neighbor
hopping ($t$) and the on-site interaction ($U$) are positive.

The Hamiltonian (\ref{hun}) is not exactly solvable by the Bethe ansatz
for $N>2$ and arbitrary $U$. It is however  possible to 
solve the generalization of the Lieb-Wu Bethe ansatz equations 
for fermions carrying a SU(N) spin index\cite{schlottman,frahm}.
The result is that for any $N>2$, there exists a Mott-Hubbard transition
from a metallic phase to an antiferromagnetic insulating phase at a 
{\it finite} value of the coupling $U$. The transition is found to be of 
{\it first}-order in contrast with the accepted view that the metal-insulator 
transition in one-dimensional systems should be of the KT type. 
The point is that a projection onto the subspace of states having at 
most two electrons at each site is crucial for the use
of the Bethe ansatz approach.  
The other configurations are automatically excluded by 
the Pauli principle in the SU(2) Hubbard model whereas for $N>2$
it is no longer the case. As a consequence, it is 
believed that the lattice model associated with the SU(N) 
generalization of Lieb-Wu Bethe ansatz equations should 
coincide with an integrable {\it non-local} version of the SU(N)
Hubbard model (\ref{hun})\cite{schlottman,frahm}. Although one naturally
expects that the true SU(N) Hubbard model will share some properties 
with its non-local partner, in particular the existence of a metallic 
phase at small enough $U$, the first-order character of the transition could take its
origin in the nonlocality of the interaction. In any case, 
in order to study (\ref{hun}) one must abandon the exact Bethe ansatz approaches and 
resort to two powerful techniques available in one dimension:
the bosonization and numerical approaches.
As we shall show, none of these techniques is by itself sufficient to
demonstrate the existence of the Mott transition. Regarding bosonization, 
the mere existence of the metal-insulator transition 
-even in the simplest scenario of a KT phase transition-
relies on the knowledge of $U$ dependence of the Luttinger parameter $K_c$,
a nonuniversal quantity which can only be computed in a perturbative expansion
in $U$. In other words, bosonization cannot tell us 
$whether$ a given lattice model will undergo a Mott-U transition. 
However, it defines a rich theoretical framework
in which many qualitative and quantitative predictions are obtained.
This provides an essential guide for the numerical investigation of a particular 
lattice model. Regarding numerical investigations the situation is also
not fully satisfactory. Beyond the evident problem of memory and CPU time
limitations, it is well-known that it is very difficult to characterize a 
KT phase transition. As we shall emphasize later, it is almost impossible to discriminate
between the opening of a charge gap at $U=0$ and at a finite positive $U$, even 
when very accurate numerical data are at our disposal. 
The strategy employed in this work will consist in combining both approaches.
Very strong evidences will be given in favor of a metal-insulator
transition occuring at a finite positive value of the interaction $U$ for
$N>2$.

Various numerical methods can be used to study the ground-state properties of 
Hamiltonian (\ref{hun}). In exact diagonalization methods \cite{dagotto} 
the exact ground-state eigenvector is calculated. 
Unfortunately, the rapid increase of the size of 
the Hilbert space restricts severely the attainable system sizes.
In order to treat bigger systems two types of approach are at our disposal: 
The density matrix renormalization group (DMRG) method and the stochastic approaches.

Since its discovery a few years ago the DMRG method has been extensively used 
for studying various one-dimensional systems
and coupled chain problems (for a review, see Ref. \cite{lectures1}, for a detailed 
presentation of the method, see Refs. \cite{white1,white2}).
DMRG is a very efficient real-space numerical 
renormalization-group (RG) approach. The fundamental point which makes the method 
successful is the way that ``important'' degrees of freedom are chosen at 
each RG iteration. Instead of keeping the lowest eigenstates of the RG block considered 
as isolated from the outside world (as it was usually done in previous approaches), the 
states which are selected are the most probable eigenstates of the density matrix 
associated with the block considered as a part of the whole system.
The main error of DMRG is related to the finite number of states kept at each 
iteration of the algorithm. In order to get the exact property the extrapolation to an
infinite number of states has to be performed. 
At least for 1D and quasi-1D problems, and for systems having a small number 
of states per site, the errors obtained are small. Note also that 
DMRG works especially well when open boundary conditions 
are used. For periodic boundary conditions, errors are significantly larger.

In this paper we use an alternative approach based on a stochastic sampling 
of the configuration space. Such approaches are referred to as 
quantum Monte Carlo (QMC) methods. There exists a large variety of QMC approaches.
A first set of methods is defined within a finite-temperature framework 
(Path-Integral Monte Carlo, World-line Monte Carlo, etc$\cdots$, see e.g. Ref. \cite{linden}).
In these approaches, the main systematic error is the high-temperature 
approximation associated with the Trotter break-up (Trotter or short-time error). 
When interested in obtaining the zero-temperature properties the 
number of ``time slices'' to consider must be taken large and the computational 
effort becomes important.  Practical calculations have shown that the method
is much less accurate than DMRG, at least for one-dimensional systems. 
In the second type of approaches used here, the stochastic sampling is directly defined  
within a zero-temperature framework. 
These methods are usually referred to as Green's function Monte Carlo 
(GFMC) or projector Monte Carlo. For systems having a nodeless ground-state
wave function as it is the case here, the GFMC method can be extremely powerful. 
The basic idea is to extract from a known trial wave function
$\psi_T$ its exact ground-state component $\psi_0$. 
To do that an operator $G({\cal H})$ acting as a filter is introduced. Statistical 
rules are defined in order to calculate stochastically the action of the operator $G$ on 
a given function. Apart from statistical fluctuations, 
the GFMC method is an exact method. 
It does not require an extrapolation to zero temperature as in finite-temperature schemes.
In addition, there exists a so-called zero-variance property
for the energy: The better the trial wave function $\psi_T$ is,
the smaller the statistical fluctuations are. In the limit of an exact
wave function, the statistical fluctuations entirely
disappear (zero-variance property). As an important consequence, 
by choosing a good enough trial wave function very 
accurate calculations can be performed (see, for example, Ref. \cite{ceperley2}). 
Note that, in contrast with DMRG, the efficiency of GFMC does not depend 
on the specific type of boundary conditions chosen and that the number of states 
per site is not a critical parameter of the simulation.
Here, it is an important point since the SU(N) model displays $2^N$ states 
per site (for the SU(4) case treated here it gives 16 states per site).

In order to improve further the accuracy of the approach we present  
a generalized version of the GFMC method in which the dynamics of the Monte Carlo 
process is partially integrated out. More precisely, we generalize an 
idea introduced by Trivedi and Ceperley in their GFMC 
study of the S=1/2 Heisenberg quantum antiferromagnet \cite{ceperley2}. In the
GFMC method the probability that the random walk remains a certain number of 
times in the same configuration is described by a Poisson distribution.  
It is then possible to sample the corresponding ``trapping time'' from this 
distribution and to weight the expectations values according to it. 
As remarked by Trivedi and Ceperley, doing this can lead to a considerable 
improvement in the simulation. This is particularly true when the wave function 
is localized (large $U$ regime for our model, systems with deep potential 
wells, etc$\cdots$). 
Here, we show that the method can be improved further by integrating out 
exactly the time evolution associated with this trapping phenomenon.
Once this is done we are left with a random walk defined by an 
``escape transition probability'' connecting non-identical configurations 
(the system never remains in the same configuration)
and a modified branching term resulting from the time-integration. 
Note that 
introducing trapping times in averages helps a lot when 
optimizing the parameters of the trial wave function. 
Finally, we present an original method for computing the Luttinger 
liquid parameters within a QMC scheme. 
We show that these parameters can be obtained from a series of 
ground-state calculations of total energies of {\it real} -but not necessarily 
Hermitian- Hamiltonians.
In this way we escape from the difficulty
of calculating with QMC ground-state energies of the {\it complex} 
Hamiltonians resulting from the definition of the charge and spin stiffnesses.
Although it is difficult to compare the efficiency of our generalized GFMC 
approach with DMRG (since the quality of GFMC simulations is too much 
dependent on the quality of the trial wave function used) 
we believe that the accuracy of our
results is comparable or even better to what can be 
done with DMRG. In any case, our data
are sufficiently accurate to conclude to the existence of a metal-insulator
phase transition in the model studied.

Very recently, Beccaria {\sl et al.} \cite{beccaria} have proposed a 
QMC algorithm based on the use of Poisson processes. Their approach 
contains similar ideas. However, in contrast with the present approach no 
importance sampling is used and no integration of the Poisson dynamics is performed.
It should also be noted that the use of Poisson processes for describing the time 
evolution of systems trapped in some configuration is not restricted to quantum systems. 
Krauth and collaborators have proposed related ideas within the context of 
classical Monte Carlo simulations \cite{krauth1,krauth2}. 

The organization of the paper is as follows.
In section II, a bosonization approach of the SU(N) Hubbard
model will be given. Some of the results has already been
obtained by Affleck\cite{affleck} whereas
additionnal new ones will also be useful to compare with 
the numerical simulations. The purpose of section III is 
to give a presentation of the GFMC method
together with our generalization based on the partial integration
of the dynamics.
The pratical implementations of 
the GFMC approach for the Hamiltonian (\ref{hun}) will be 
discussed in section IV and the numerical results for $N=2,3,4$ 
will be presented in section V. Finally, section VI gives a summary
of the work together with a comparison between the physical 
results obtained for the SU(N) Hubbard model and those corresponding 
to its nonlocal integrable version. In the
Appendix we give some details of computation occuring in section II.  

\section{The SU(N) Hubbard Model}

In this section, we shall use a bosonization approach (for recent
reviews see Refs. \cite{voit,tsvelik})
to study the SU(N) Hubbard model.  
Before doing that, let us first discuss the symmetries of the model. 

The Hamiltonian (\ref{hun}) has a U(1)$\otimes$ SU(N) symmetry:
\bea
c_{ia} &\rightarrow& \re^{i \theta} c_{ia} \nonumber \\
c_{ia} &\rightarrow& U_{ab} c_{ib} 
\label{U1SUN}
\eea
where the matrix $U$ belongs to SU(N).
These symmetries express the
conservation of the charge and spin invariance under a SU(N) rotation.
The associated generators are given by the 
following operators:
\bea
{\cal N} &=& \sum_{i,a} n_{ia}  \nonumber \\
{\cal S}^A &=&\sum_{i} {\cal S}^A_i
\eea
with
\be
{\cal S}^A_i = c_{ia}^{\dagger} { {\cal T}_{ab}^{A} } c_{ib}
\label{sunspin}
\ee
where the summation over repeated indexes (except for lattice
indexes)
is assumed in the following. 
In the latter equation, the  N$^2$ - 1 matrices ${\cal T}^A$ are the generators
of the Lie algebra of SU(N) in the fundamental representation. They
satisfy the commutation relation:
\be
[{\cal T}^A, {\cal T}^B] = i f^{ABC} {\cal T}^C,
\label{commulie}
\ee
$f^{ABC}$ being the structure constants of the Lie algebra
and the generators 
are normalized according to Tr(${\cal T}^A {\cal T}^B$) = $\delta^{AB}/2$.
The conservation law associated with the U(1) symmetry allows to study
the Hamiltonian (\ref{hun}) for a fixed density $n$. 
The Coulomb interaction can thus be rewritten, up to a constant,  
in terms of the SU(N) spin operator:
\be
\frac{U}{2} \;  \left(\sum_{a=1}^N n_{ia}\right)^2 =
- \frac{UN}{N+1} {\cal S}^A_{i} {\cal S}^A_{i}
\label{intspin}
\ee
where we have used the identity: 
\be 
{\cal T}^A_{ab} {\cal T}^A_{de}
= \frac{1}{2}\left(\delta_{ae} \delta_{bd} - \frac{1}{N} 
\delta_{ab} \delta_{de}\right). 
\label{geniden}
\ee
The relation (\ref{intspin})
makes explicit the SU(N) invariance of the model.

The Hamiltonian (\ref{hun}) is not exactly solvable by 
the Bethe ansatz for N$>$2 and arbitrary $U$, even if, as already emphasized, 
some integrable non-local extension 
of (\ref{hun}) with a SU(N) symmetry can be considered.
The situation is simpler in the limit $ U \rightarrow \infty$
and at half filling (one ``electron'' per
site), i.e. when $k_F = \pi/Na_0$ ($a_0$ 
being the lattice spacing). 
In that case, 
it can be shown that
(\ref{hun}) reduces to the SU(N) Heisenberg antiferromagnetic chain for which
an exact solution is available. 
As shown by Sutherland\cite{sutherland}, this latter model is critical with 
$N-1$ massless bosonic modes with the same velocity.
In the Conformal Field Theory (CFT) language, 
the central charge of the model in the infrared (IR) limit is 
$c=N-1$ and using a non-Abelian bosonization of (\ref{hun}),
Affleck\cite{affleck} identifies the nature of the critical theory in the 
spin sector as the SU(N)$_1$ Wess-Zumino-Novikov-Witten (WZNW)
model. In the following, we shall present both non-Abelian and Abelian bosonization 
approaches of the SU(N) Hubbard model (\ref{hun}) at half filling and give
a number of results that will be essential for discussing the numerical data
presented in section V.

\subsection{ Continuum limit}

In the continuum limit, 
the spectrum around the two Fermi points $\pm k_F$
is linearized and gives rise to left-moving fermions $\psi_{aL}$
and right-moving fermions $\psi_{aR}$. In this low-energy procedure, 
the lattice fermion operators $c_{ia}$ are expressed in terms of 
these left-right moving fermions as:
\bea
\frac{c_{ia}}{\sqrt{a_0}} \rightarrow \psi_a\left(x\right) \sim
\psi_{aR}\left(x\right) \re^{i k_F x} + 
\psi_{aL}\left(x\right) \re^{-i k_F x},
\label{ccont}
\eea
where  $x = ia_0$.  
In this continuum limit, the non-interacting part of the 
Hamiltonian (\ref{hun})
corresponds to the  Hamiltonian density 
of $N$ free relativistic fermions: 
\be
{\cal H}_0 = - i v_F \left(:\psi^{\dagger}_{aR}\partial_x \psi_{aR}:
- :\psi^{\dagger}_{aL}\partial_x \psi_{aL}: \right)
\label{h0}
\ee
where 
the normal ordering $::$ with 
respect of the Fermi sea is assumed and the Fermi velocity $v_F$ is 
given by:
\be
v_F = 2t a_0 \sin \frac{\pi}{N}.
\label{velofermi}
\ee
In the continuum limit, the SU(N) spin operator (\ref{sunspin})
decomposes into an uniform and a $2k_F$ contribution:
\bea
\frac{{\cal S}^A_i}{a_0} \rightarrow {\cal S}^A\left(x\right) 
\simeq {\cal J}^A\left(x\right) + \left(\re^{2ik_F x} {\cal N}^A\left(x\right) 
+ {\rm H.c.} \right)
\label{spincont}
\eea
where the $2k_F$ contribution is given by:
\be
{\cal N}^A = 
\psi^{\dagger}_{aL}{\cal T}^A_{ab}\psi_{bR}
\label{2kfcurrfer}
\ee
whereas the uniform part reads:
${\cal J}^A = {\cal J}^A_R + {\cal J}^A_L$ with 
\be
{\cal J}^A_{R(L)} = 
:\psi^{\dagger}_{aR(L)}{\cal T}^A_{ab}\psi_{bR(L)}:.
\label{currfer}
\ee
These left-right SU(N) spin currents obey the following
Operator Product Expansion (OPE) (see the Appendix):
\bea
\lim_{x \rightarrow y}{\cal J}^A_{R(L)}\left(x\right) 
{\cal J}^B_{R(L)}\left(y\right) \sim 
\frac{-  \delta^{AB}}{8\pi^2 \left(x - y\right)^2} 
\nonumber \\
\mp \frac{f^{ABC}}{2\pi \left(x - y\right)}
{\cal J}^C_{R(L)}\left(y\right)
\label{OPEspin}
\eea
which shows that they satisfy the SU(N)$_1$ Kac-Moody (KM) 
algebra\cite{tsvelik,difrancesco}. 
In the same way, the total charge density $\sum_a n_{ia}$ reads 
in the continuum limit:
\be
\sum_a n_{ia} \rightarrow a_0^{1/2} \left(
{\cal J}^0 \left(x\right) + 
\left(\re^{-2ik_F x} \psi^{\dagger}_{aR}\left(x\right) 
\psi_{aL}\left(x\right)
 + {\rm H.c.} \right) \right)
\ee
where ${\cal J}^0 = {\cal J}^0_R + {\cal J}^0_L$ and
\be
{\cal J}^0_{R(L)} =  
:\psi^{\dagger}_{aR(L)}\psi_{aR(L)}:
\label{chargedens}
\ee
are the U(1) right and left charge currents. 
These currents satisfy the OPE: 
\be 
\lim_{x \rightarrow y}{\cal J}^0_{R(L)}\left(x\right) 
{\cal J}^0_{R(L)}\left(y\right) \sim 
-\frac{N}{4\pi^2 \left(x - y\right)^2} 
\label{OPEcharge}
\ee
and ${\cal J}^0_{R(L)}$ belongs to the U(1)$_N$ KM algebra.

With these identifications, it is not 
difficult to show (see the Appendix) that the free
part of the Hamiltonian (\ref{h0}) can be expressed only in terms of spin and
charge currents (the so-called Sugawara form):
\be
{\cal H}_0 = {\cal H}_{0s} + {\cal H}_{0c}
\ee
with
\be
{\cal H}_{0s} = \frac{2\pi v_F}{N+1} \; 
\left(:{\cal J}_R^A {\cal J}_R^A: + :{\cal J}_L^A {\cal J}_L^A:
\right) 
\label{sugawaraspin}
\ee
and 
\be
{\cal H}_{0c} = \frac{\pi v_F}{N} 
\left(:{\cal J}_R^{0}{\cal J}_R^{0}: + :{\cal J}_L^{0} {\cal J}_L^{0}: 
\right).
\label{sugawaracharge}
\ee
At the level of the free theory,
spin and charge degrees of freedom decouple.
The symmetry of the free Hamiltonian ${\cal H}_0$ 
in the continuum limit is therefore enlarged to give:
U(1)$_L \otimes$ SU(N)$_L \otimes$ 
U(1)$_R \otimes$ SU(N)$_R$. 
The Hamiltonian ${\cal H}_{0s}$ is nothing but the 
Sugawara form of the SU(N)$_1$ WZNW 
model\cite{tsvelik,difrancesco}.  
It is a conformaly invariant theory with central charge $c=N-1$
($N-1$ massless bosons).  The contribution ${\cal H}_{0c}$
describes the U(1) charge degrees of freedom and has central charge
$c=1$ ($1$ massless boson). 

The non-trivial part of the problem stems
from the Coulomb interaction
(\ref{intspin}). At sufficiently small $U <<$t, from Eq. (\ref{intspin}), 
we see that its contribution will be given
by the OPE:
\be
{\cal V}\left(x\right) = - U a_0 
\; \frac{N}{N+1}  \lim_{\epsilon \rightarrow 0}
{\cal S}^A\left(x+\epsilon\right){\cal S}^A\left(x\right).
\ee
From Eq. (\ref{spincont}),  there are three contributions to 
${\cal V}$:
\be
 {\cal V} = {\cal V}_{0} + {\cal V}_{2k_F} + {\cal V}_{4k_F}.
\ee
The first term is the uniform $k=0$ component while the others
contain oscillating factors $\re^{\pm 2ik_F x}$ and 
$\re^{\pm 4ik_F x}$. Neglecting all oscillatory contributions,
we are thus left with the uniform part ${\cal V}_{0}$.
Performing the necessary OPEs (see 
the Appendix), one finds that the total effective low energy
Hamiltonian density separates into two commuting charge and spin
parts:
\be
{\cal H}= {\cal H}_c + {\cal H}_s
\label{spinchargesepar}
\ee
with
\be
{\cal H}_c = \frac{\pi v_c}{N} \left(:{\cal J}^0_{R}{\cal J}^0_{R}: 
+ :{\cal J}^0_{L} {\cal J}^0_{L}: \right)
+ G_c \; {\cal J}^0_{R}{\cal J}^0_{L}
\label{hcharge}
\ee
and 
\be
{\cal H}_s = \frac{2\pi v_s}{N+1} \; 
\left(:{\cal J}_{R}^A {\cal J}_{R}^A: + :{\cal J}_{L}^A {\cal J}_{L}^A:
\right) + G_s \; {\cal J}_{R}^A{\cal J}_{L}^A
\label{hspin}
\ee
where the renormalized velocities are:
\bea
v_s &=& v_F - \frac{Ua_0}{2\pi} \nonumber \\
v_c &=& v_F + (N-1)\frac{Ua_0}{2\pi} 
\label{velo}
\eea
and the current-current couplings in the charge and 
the spin sectors are given by:
\bea
G_c &=& \frac{N-1}{N} \; U a_0\nonumber \\
G_s &=& - 2 U a_0 .
\label{coupl}
\eea
Apart from a velocity renormalization, the effect of the Coulomb interaction
is exhausted in the two marginal interactions in both
charge and spin sectors. When $U>0$, the spin current-current interaction
is  marginal irrelevant. 
At the IR fixed point, G$_s^*$ = 0, the Hamiltonian in the spin sector
is that of the  SU(N)$_1$ WZNW model.
On the other hand, the current-current interaction
in the charge sector is exactly marginal since on can diagonalize
${\cal H}_c$ with a Bogolioubov transformation to recover 
the Tomonaga-Luttinger Hamiltonian. Therefore,   
${\cal H}_c$ describes the line of
fixed points of the Luttinger liquid. 

From the above analysis we 
conclude that the SU(N) Hubbard model at 
half filling
is massless for small $U>0$. The spin sector is described by the
SU(N)$_1$ WZNW model 
while the charge sector is a Luttinger liquid with continuously
varying exponents. The main point in the above analysis is the absence
of umklapp terms which, when $N=$2, opens a gap
in the charge sector for an infinitesimal 
value of the interaction. 
At this point it is worth recalling that
 the main approximation made in the above analysis is
 the omission of the oscillating contributions
${\cal V}_{2k_F}$ and ${\cal V}_{4k_F}$. This  is  a reasonable 
assumption as far as $U$ is not too large. However one expects, on general
grounds, that umklapp processes should contribute at sufficiently large $U$ and
that a Mott transition to an insulating phase should occur at a finite
$U_c$. Indeed, in the $U\rightarrow \infty$ limit,
we have an insulating phase where the spin degrees of freedom 
are described by the SU(N) Heisenberg antiferromagnet.
We shall return to this point later. For now let us concentrate
on the properties of the metallic phase.

\subsection{The metallic phase}

At this point, we introduce $N$ chiral bosonic fields
$\phi_{aR(L)}$, $a=(1,...,N)$, using the Abelian 
bosonization of Dirac fermions\cite{tsvelik}:
\be
\psi_{aR(L)} = \frac{\kappa_{a}}{\sqrt{2\pi}}
:\exp\left(\pm\ri \sqrt{4\pi} \phi_{a R(L)}\right): 
\label{bosofer}
\ee
where the bosonic fields satisfy the following commutation 
relation $\left[\phi_{a R}, \phi_{b L}\right] = \frac{i}{4} 
\delta_{ab} $.
The anticommutation between fermions with different spin indexes is
realized through the presence of Klein factors (here Majorana fermions) $\kappa_{a}$
with the following anticommutation rule: $\{\kappa_{a}, \kappa_{b}\} = 
2 \delta_{ab}$. As in the $N=2$ case, it is suitable to switch to a basis where
the charge and spin degrees of freedom single out. 
To this end, let us introduce a charge bosonic field $\Phi_{cR(L)}$ and $N-1$
spin bosonic fields $\Phi_{msR(L)}$, $m=(1,...,N-1)$ as follows:
\bea
\Phi_{cR(L)} &=&\frac{1}{\sqrt N} \left(\phi_{1} + ...+ \phi_{N} \right)_{R(L)}
 \nonumber \\
\Phi_{msR(L)} &=&\frac{1}{\sqrt{m(m+1)}}\left( \phi_{1} + ...+\phi_{m}
- m \phi_{m+1}\right)_{R(L)} .
\label{can}
\eea
The transformation (\ref{can}) is canonical and preserves the bosonic 
commutation relations.
The inverse transformation is easily found to be:
\bea
\phi_{1R(L)}= \frac{1}{\sqrt N}\Phi_{cR(L)} + \sum_{l=1}^{N-1}
\frac{\Phi_{lsR(L)}}{\sqrt{l(l+1)}} \nonumber \\
\phi_{aR(L)}= \frac{1}{\sqrt N}\Phi_{cR(L)}   
- \sqrt{\frac{a-1}{a}} \Phi_{(a-1)sR(L)}\nonumber \\
+ \sum_{l=a}^{N-1}
\frac{\Phi_{lsR(L)}}{\sqrt{l(l+1)}}, \;\;
a=2,...,N-1 \nonumber \\
\phi_{NR(L)}= \frac{1}{\sqrt N}\Phi_{cR(L)} 
- \sqrt{\frac{N-1}{N}} \Phi_{(N-1)sR(L)}.
\label{caninv}
\eea
In this new basis, the Hamiltonian density in the 
spin sector at the SU(N)$_1$ fixed point
reads:
\be
{\cal H}^*_s 
= \frac{u_s}{2} \; \sum_{m=1}^{N-1}
\left(:\left(\partial_x\Phi_{ms}\right)^2 :+
 :\left(\partial_x\Theta_{ms}\right)^2:\right)
\label{hspinboso}
\ee
where $u_s$ is the spin velocity at the fixed point and
\bea
\Phi_{ms} &=& \Phi_{msL} + \Phi_{msR} \nonumber \\
\Theta_{ms} &=&\Phi_{msL} - \Phi_{msR}.
\eea
This representation makes
clear the fact that the central charge in the spin sector is 
indeed $c=N-1$. 

Let us now concentrate
on the charge sector. 
It is not difficult to show, using Eqs. (\ref{chargedens}, \ref{bosofer}) 
and (\ref{can})
that the charge current expresses as:
\be
{\cal J}^0_{R(L)} = \sqrt {\frac{N}{\pi}} \partial_x\Phi_{cR(L)} 
\label{chargecurr}.
\ee
Therefore, the Hamiltonian density (\ref{hcharge}) in the charge sector reads:
\bea
{\cal H}_c = \frac{v_c}{2} \;\left(:\left(\partial_x\Phi_{c}\right)^2 : +
 :\left(\partial_x\Theta_{c}\right)^2:\right)  \nonumber \\
+ (N-1) \frac{U a_0}{\pi} \;  \partial_x\Phi_{cL}\partial_x\Phi_{cR}
\label{hchargeboso}
\eea
where we have introduced the total charge bosonic field: 
$\Phi_c = \Phi_{cR} + \Phi_{cL}$ and its dual: 
$\Theta_c = \Phi_{cL} - \Phi_{cR}$.
The Hamiltonian (\ref{hchargeboso})
can be written in the Luttinger liquid form:
\be
{\cal H}_c = \frac{u_c}{2}\;\left( \frac{1}{K_c} 
:\left(\partial_x\Phi_{c}\right)^2 :+ \; 
K_c \; :\left(\partial_x\Theta_{c}\right)^2:\right)
\label{hlutt}
\ee
where the charge exponent $K_c$ and the 
renormalized charge velocity $u_c$ are given by:
\bea
K_c &=& \frac{1}{ \sqrt{1  + (N-1)U a_0/\pi v_F} } \nonumber \\
u_c  &=& v_F \sqrt{1  + (N-1)Ua_0/\pi v_F} .
\label{luttpara}
\eea
 The $U$ dependence of the Luttinger
parameters $K_c$ and $u_c$ given in the 
above expressions should not be taken too seriously. Indeed, 
the continuum limit
approach is strictly speaking valid only provided $U/t<<1$. In this regime 
one has
\bea
K_c &\sim & 1 - (N-1) \frac{U a_0}{2\pi v_F} \nonumber \\
u_c &\sim& v_F + (N-1)\frac{U a_0}{2\pi}.
\label{luttparaper}
\eea
The physically relevant question is now what happens for higher values
of the interaction $U$. In the absence of umklapp terms, the accepted view
is that the effect of interaction corresponds to a 
renormalization of the Luttinger
parameters $K_c$ and $u_c$ as well as the spin velocity $u_s$
which have therefore to be thought as phenomenological
parameters as the Landau coefficients in the Fermi 
liquid theory\cite{haldane,voit}. 
These parameters
completely characterize the low energy properties of the metallic phase
as we shall see now. Let us first discuss the  
electronic Green's function defined by:
\be
G_{ab}(x,\tau) = \langle  \psi_a^{\dagger}(x,\tau) \psi_b(0,0) \rangle ,
\ee
$\tau$ being the imaginary time.
This correlation function 
can be computed using Eqs. (\ref{ccont}, \ref{bosofer}) and (\ref{caninv}). 
After some calculations, one finds:
\bea
G_{ab}(x,\tau) &\sim& \frac{\delta_{ab}}{2\pi} 
\left[\frac{1}{x^2 + u_c^2 \tau^2}\right]^{\alpha/2}
\nonumber \\
&\times& 
 \left[ \frac{\exp\left(ik_F x\right)}{\left(ix + u_c \tau\right)^{1/N}
\left(ix + u_s \tau\right)^{1-1/N} }
\right.\nonumber \\
&+& \left. \frac{\exp\left(-ik_F x\right)}{\left(-ix + u_c \tau\right)^{1/N}
\left(-ix + u_s \tau\right)^{1-1/N} } \right]
\label{greenelec}
\eea
where the exponent $\alpha$ is given by:
\be
\alpha = \frac{1}{2NK_c} \; \left(1 - K_c\right)^2 .
\label{alpha}
\ee
This allows to give an estimate of the single particle density of states
which is related to the electronic Green's function at $x=0$:
\be
\rho(\omega) \sim  |\omega|^{\alpha}.
\label{densstates}
\ee
Similarly, $K_c$ determines the singularity of the momentum 
distribution $n_a(k)$ around the Fermi point $k_F$: 
\be 
n_a\left(k\right) = n_a\left(k_F\right) + {\rm Cte} \;\;{\rm sgn} 
\left(k-k_F\right) 
|k-k_F|^{\alpha}
\ee
and the momentum distribution function has a power law 
singularity at the Fermi level unlike a standard Fermi liquid.
This anomalous power law behavior for any finite value of $N$ 
is inherent of a Luttinger liquid. 

The computation of the SU(N) spin-spin correlation function:
\be
\Delta^{AB}(x,\tau) = \langle{\cal S}^A\left(x,\tau\right)
{\cal S}^B\left(0,0\right)\rangle
\ee
is more involved. It can be shown that the 
leading asymptotics of this correlation function is
given by the $2k_F$ part:
\be
\Delta^{AB}(x,\tau) \sim
\delta^{AB} \frac{\cos\left(2k_Fx\right)}
{\left(x^2+u_c^2\tau^2\right)^{K_c/N}\left(x^2+u_s^2\tau^2\right)^{1-1/N}}.
\label{greenspin}
\ee
We deduce from the above 
correlation function the low temperature dependence of the
NMR relaxation rate T$_1$:
\bea
\frac{1}{T_1 T} &\sim& T^{2/N + 2K_c/N -2}.
\label{nmr}
\eea

As seen, once the $U$ dependence of the Luttinger parameters
$u_c$, $K_c$ and $u_s$ is known, the low energy properties of 
the metallic phase are
entirely determined.  These parameters are non-universal and cannot 
be obtained for arbitrary $U$
by the continuum limit approach. Although $K_c < 1$ when $U>0$, one does not 
know its minimum value. It is only in the $N=2$ case, 
that the Luttinger parameters can be extracted from 
the exact solution\cite{schulz1,korepin,kawakami}.
When $N>$2 no exact solution is available and one has to use 
numerical computations to estimate
these parameters. This will be done for the two cases $N=3$ and $N=4$
in section V. Before doing that, let us discuss about the Mott transtion
that should occur in the 
problem for a finite critical value of the repulsion $U$ for
$N >2$.

\subsection{ The Mott transition}

The very difference between the $N=2$ and $N>2$ cases lies in the fact that
there is no umklapp term at half filling in the bare Hamiltonian in the 
continuum limit. 
The reason for this is that
these terms came with oscillating factors and were 
omitted for small value of the repulsion. 
However, in  the RG strategy one has 
to look at the stability of the Luttinger
fixed line and any operator that is compatible with the symmetry of the problem
should be taken into account: they will be generated during the 
renormalization process.
In our problem, the important symmetries  are the SU(N)
spin rotation invariance, chiral invariance and translation invariance. 

From Eqs. (\ref{ccont}, \ref{bosofer}) and (\ref{can}), one easily finds 
that under a translation by one lattice site, the charge field $\Phi_{c}$ is
shifted according to:
\be
\Phi_{c} \rightarrow \Phi_{c} + \sqrt{\frac{\pi}{N}}.
\label{shift}
\ee
Therefore one can add any operator in the charge sector 
that is invariant under the transformation (\ref{shift})
and will be necessary generated by higher order in 
perturbation theory. The operator with
the smallest scaling dimension that is invariant under (\ref{shift}) is:
\be
{\cal H}_{umklapp} = - G_u :\cos\left(\sqrt{4\pi N}\Phi_{c}\right):.
\label{umklapp}
\ee
Other operators, with higher scaling dimensions,
that couple spin and charge degrees of freedom may also be included. This
is the reason why one cannot exclude the possibility of a charge density 
wave (CDW) instability. For instance, such processes are present in 
the extended SU(2) Hubbard model at half filling\cite{voit1}.
Alhough it requires some formal proof, 
we expect that, due to the fact that in the present model
the interaction is local in the density, 
the leading umklapp contribution should only 
affect the charge sector. We have checked that this is indeed 
true for the particular cases, $N=3$ and $N=4$\cite{azaria}.
We have shown indeed by perturbation theory that the oscillating
contributions
${\cal V}_{2k_F}$ and ${\cal V}_{4k_F}$
generate $6k_F$ and $4k_F$ processes for $N=3$ and $N=4$ respectively.
Up to irrelevant operators, the only contribution we found is precisely
(\ref{umklapp}) with $N=3$ and $N=4$. 
In any case in what follows, we shall thus make the
hypothesis, first made by Affleck\cite{affleck},
that all the effects of high energy processes are exhausted
by (\ref{umklapp}) for the general SU(N) case. Consequently,  
the effective Hamiltonian density in the spin sector is still given
by the SU(N)$_1$ WZNW model and the effective
Hamiltonian in the charge sector is now:
\bea
{\cal H}_c = \frac{u_c}{2}\;\left(\frac{1}{K_c} 
:\left(\partial_x\Phi_{c}\right)^2 :+ \; 
K_c \; :\left(\partial_x\Theta_{c}\right)^2:\right)  \nonumber \\
- G_u :\cos\left(\sqrt{4\pi N}\Phi_{c}\right): .
\label{hmott}
\eea
Rescaling the fields as $ \Phi_{c} \rightarrow \Phi_{c}\sqrt{ K_c}$
and $\Theta_{c} \rightarrow \Theta_{c}/\sqrt{K_c}$, 
the Hamiltonian (\ref{hmott}) in the charge sector becomes
the Hamiltonian of the sine-Gordon model:
\bea
{\cal H}_c = \frac{u_c}{2}\;\left(:\left(\partial_x\Phi_{c}\right)^2 :+
\; :\left(\partial_x\Theta_{c}\right)^2:\right)  \nonumber \\
- G_u :\cos\left(\sqrt{4\pi N K_c}\Phi_{c}\right): .
\label{hmottsg}
\eea
Since the scaling dimension of the cosine term 
in (\ref{hmottsg}) is $\Delta_u = N K_c$,
we deduce that a gap opens in the charge sector when:
\be
K_c = \frac{2}{N}. 
\label{Kcrt}
\ee 
On the other hand, when $K_c < 2/N$, the umklapp term is irrelevant 
and the system
remains in the metallic phase described in 
the preceeding subsection.
Therefore, as $U$ increases,  $K_c$ will decrease from $1$ at $U=0$
to $K_c = 2/N$  at a critical value of the interaction $U_c$
where a Mott transition to an insulating phase occurs. Within this scheme,
the phase transition is expected to be of the KT type. Of course
when $U>U_c$,  $K_c$ vanishes so that the jump 
is $1 - 2/N$ and is universal. The present approach cannot give an 
accurate value of $U_c$. However, one can get a rough estimate
of $U_c$ using Eqs. (\ref{velofermi}, \ref{luttpara}) and (\ref{Kcrt}):
\be
\frac{U_c}{t} = \frac{\pi}{2} \; \frac{N^2 - 4}{N-1} \; \sin\frac{\pi}{N}.
\label{ucrt}
\ee

In the insulating phase, the charge field $\Phi_{c}$ is locked 
in a special well of the sine-Gordon model (\ref{hmottsg}) 
and the leading asymptotics of the SU(N) spin-spin correlation 
functions is now: 
\be
\Delta^{AB}(x,\tau) \sim \lambda_1
\delta^{AB} \frac{\cos\left(2k_Fx\right)}{\left(x^2+u_s^2\tau^2\right)^{1-1/N}}
\label{greenspinmott}
\ee
where $\lambda_1$ is a non-universal constant stemming from the 
charge degrees of freedom. One recovers the result 
previously derived by Affleck\cite{affleck}.
The NMR relaxation rate 
behaves now as 1/(T$_1$T $) \sim$ T$^{2/N - 2}$. Finally, 
let us note that there are other harmonics $4k_F, 6k_F..$ 
in the SU(N) spin density (\ref{spincont}) that will be 
generated by higher orders in perturbation theory. Together with 
the uniform contribution with scaling dimension $1$, these terms 
will give subleading power law contributions in the SU(N) spin-spin 
correlation function (\ref{greenspinmott}). These operators correspond to the
primary fields of SU(N)$_1$ WZNW transforming to other representation
of SU(N) than the fundamental one. 
One should recall that for the SU(N)$_1$ WZNW, 
there are $N-1$ primary fields\cite{difrancesco}.
A primary field ${\tilde \phi}_a$ ($a=1,..,N-1$)
of SU(N)$_1$ transforms according to the a$^{th}$
basic representation of SU(N) (Young tableau 
with $a$ boxes and a single column) and has scaling
dimension: $\Delta_a = a(N-a)/N$.
We thus expect the following asymptotics for 
$\Delta^{AB}$ with some non-universal constants ($\lambda_a$): 
\bea 
\Delta^{AB}(x,\tau) \sim -\frac{\delta^{AB}}{8\pi^2} \left( 
\frac{1}{\left(u_s \tau - \ri x\right)^2} + 
\frac{1}{\left(u_s \tau + \ri x\right)^2} \right) \nonumber \\
+ \delta^{AB} \sum_{a=1}^{N-1} \lambda_a
\frac{\cos\left(2 a k_F x\right)}
{\left(x^2+u_s^2\tau^2\right)^{a-a^2/N}}
\label{greenspinmottcomp}
\eea
up to logarithmic contributions originating from the marginal
irrelevant current-current interaction in the spin sector\cite{itoi}. 

We end this subsection by giving 
the low-temperature expression of the uniform susceptibility $\chi$ and 
the specific heat  of the SU(N) Hubbard model in the insulating antiferromagnetic phase.
The continuum density that describes the behavior of the SU(N) spins
degrees of freedom in a uniform magnetic field $H$ is given by:
\bea 
{\cal H}_H = 
\frac{u_s}{2} \; \sum_{m=1}^{N-1}
\left(:\left(\partial_x\Phi_{ms}\right)^2 :+
 :\left(\partial_x\Theta_{ms}\right)^2:\right) \nonumber \\
- H \sum_{m=1}^{N-1} {\cal J}^m 
\label{hmag}
\eea
where we have neglected the marginally irrelevant current-current interaction.
In Eq. (\ref{hmag}), we have considered a uniform magnetic field 
along the diagonal ${\cal T}^m, (m=1,..,N-1)$ generators of SU(N)
that span the Cartan subalgebra of SU(N). 
According to our normalization convention, they can be written in
$N\times N$ diagonal matrices as follows: 
\bea 
{\cal T}^m = \frac{1}{\sqrt{2m\left(m+1\right)}} \;
{\rm diag}\left(1,1,...,-m,0,..,0\right)
\label{diaggen}
\eea
with $m=1,..,N-1$ and
$-m$ is located on the $m+1$ element of the diagonal. 
Using the bosonization correspondence (\ref{bosofer})
and the canonical transformation (\ref{caninv}),
the total density Hamiltonian (\ref{hmag}) in a magnetic 
field can be written as:
\bea
{\cal H}_H =
\frac{u_s}{2} \; \sum_{m=1}^{N-1}
\left(:\left(\partial_x\Phi_{ms}\right)^2 :+
 :\left(\partial_x\Theta_{ms}\right)^2:\right) \nonumber \\
- \frac{H}{\sqrt{2\pi}} \sum_{m=1}^{N-1} \partial_x \Phi_{ms}. 
\label{hmagfin}
\eea
Doing the substitution: 
\be 
\partial_x \Phi_{ms} \rightarrow 
\partial_x \Phi_{ms} + \frac{H}{\sqrt{2\pi} u_s},
\ee
we obtain the expression of the uniform susceptibility
of the SU(N) Heisenberg antiferromagnet: 
\be 
\chi = \frac{N-1}{2\pi u_s}
\label{suscept}
\ee
which is nothing but $N-1$ times the uniform susceptibility
of the SU(2) Heisenberg antiferromagnet. This result is easy to understand
since the critical theory in the spin sector corresponds to
$N-1$ decoupled massless bosonic modes.
Finally, using the general formula of the specific heat at low temperatures
for a conformaly invariant theory\cite{cardy}, one 
has for the SU(N) Heisenberg antiferromagnet:
\be 
C_V = \frac{\pi\left(N-1\right)}{3 u_s} T.
\label{specific}
\ee

Before closing this section, it is important to emphasize 
that the Mott transition expected in the bosonization approach 
relies on  the full expression of $K_c(U)$ as function of the  
interaction. However, one should stress 
that this parameter cannot be obtained for arbitrary $U$
within this approach and only in the weak coupling limit $U \ll t$ where 
the model is in its metallic phase. 
To conclude in favour of the existence of a  Mott transition 
for a finite 
value of the Coulomb interaction, one has thus to compute $K_c(U)$ 
of the lattice model by an independent approach. 
Since the SU(N) Hubbard model with $N>2$ is not exactly
soluble, one cannot determine the expression  
$K_c(U)$ by the Bethe ansatz as for 
the standard Hubbard model\cite{schulz1,korepin,kawakami}.
We shall thus compute the value $K_c(U)$ 
of the lattice model using very acurate numerical calculations
based on QMC methods described in the next section.
In section V,
we shall then compare the numerical results with the predictions of the 
bosonization approach given in this section to conclude 
on the existence of a Mott transition in the model.

\section{The numerical approach}

In this section we present our improved zero-temperature Green's function Monte 
Carlo method used for computing ground-state properties. In the first 
part a sketchy but self-contained presentation of the basic GFMC
method is given. In addition to introducing our notations for the next part, 
this section will enable the interested reader to 
understand all the practical aspects of the method.
The second part is devoted to the presentation of the generalized GFMC
method itself.

\subsection{Green's function Monte Carlo}

As already noticed in the introduction the basic idea of the GFMC 
method is to extract from a known trial wave function 
$|\psi_T>$ the exact ground-state component $|\psi_0>$. 
To do that an operator $G({\cal H})$ acting as a filter is introduced. 
For continuum problems standard choices are $G({\cal H})=
\exp(-\tau {\cal H})$ 
(Diffusion Monte Carlo) or $G({\cal H})= 1/(1+\tau(E-{\cal H}))$ (Green's
function Monte Carlo). 
For a lattice problem or any model with a finite number of 
states (finite matrix)
a natural choice to consider is 
\be
G({\cal H}) \equiv  1-\tau({\cal H}-E_T)
\label{proj2}
\ee
where $\tau$ plays the role of a time-step (a positive constant) 
and $E_T$ is some reference energy.
If $\tau$ is chosen sufficiently small and $|\psi_T>$
has a non-zero overlap with the ground-state, the exact ground-state is 
filtered out as follows
\be
\lim_{P \rightarrow \infty} {G({\cal H})}^P|\psi_T> \sim |\psi_0>.
\label{lim2}
\ee
This result is easily obtained by expanding $|\psi_T>$ within the complete set 
of eigenstates of ${\cal H}$.

In Monte Carlo schemes, successive applications 
of the operator $G({\cal H})$ 
on $|\psi_T>$ are done using probabilistic rules. These rules 
are implemented in configuration space where the trial wave function and 
matrix elements of ${\cal H}$ are easily evaluated. In what 
follows we shall denote by
$\mid i\rangle$ an arbitrary configuration of the system. To give an example, 
in actual calculations presented below we consider 
$\mid i \rangle = |i^{(1)}>\cdots|i^{(N)}>$ with $|i^{(a)}> \equiv
| n_{1 a},\cdots,n_{L a}>$
where $L$ is the number of sites, $a$ the SU(N) color index, 
and $n_{i a}$ the occupation number of site $i$ ($n_{i a}=0$ or $1$)
for the species $a$.

In this work Hamiltonians considered are of the form: 
\be
{\cal H}=T+V
\label{H}
\ee
where $T$ is the kinetic term (a non-diagonal operator) and
$V$ is a (diagonal) potential term. For fermions in one dimension 
it is known that by choosing a suitable labelling 
of the sites, matrix elements of the kinetic term can all be made negative
\be
<i|T|j> \;\;  < 0  \;\;\;\; \;\;  (i \ne j).
\label{tij}
\ee
A most important consequence of this property is that the exact ground-state 
has a constant sign. In other words, simulations presented here are free of 
the sign problem.

Let us now introduce the following transition probability 
\be
P_{i \to j}(\tau) = \psi_T(j) \langle j 
\mid [1- \tau ({\cal H}-E_L)]\mid i \rangle
\frac{1}{\psi_T(i)} 
\label{pij}
\end{equation}
where $\psi_T(i)$ are the components of 
the vector $|\psi_T>$, $\psi_T(i) \equiv <i|\psi_T>$,
and $E_L$ is a diagonal operator called the local energy and defined as follows
$$
<i|E_L|j>=\delta_{ij} E_L(i)
$$
with
\be
E_L(i)= \frac{<i|{\cal H}|\psi_T>}{<i|\psi_T>}.
\label{el}
\ee
Note the important relation associated with the definition of the local 
energy:
\begin{equation}
({\cal H}-E_L)\mid \psi_T\rangle=0.
\label{relEL}
\end{equation}
To define a transition probability $P_{i \to j}$ must fulfill 
the two following conditions.
First, the sum over final states, $\sum_j P_{i \to j}$, must be equal to one. 
Here, this is true as a direct consequence of (\ref{relEL}).
Second, $P_{i \to j}$ must be positive. To see this and for later use, 
let us distinguish between the cases $i=j$ 
and $i \ne j$. 

For $i=j$ we have
\begin{equation}
P_{i \to i}(\tau)= 1 +\tau T_L(i)   
\label{pii}
\end{equation}
where $T_L(i) \equiv E_L(i)-H_{ii}$. Using (\ref{H}), $T_L(i)$ can 
be rewritten as
\begin{equation}
T_L(i)= \frac{<i|T|\psi_T>}{<i|\psi_T>},
\label{tli}
\end{equation}
$T_L(i)$ is called the local kinetic energy. Because of (\ref{tij}) 
it is a negative quantity and the transition probability can be made positive 
by taking $\tau$ sufficiently small.  More precisely, the time-step must verify
\begin{equation}
0 < \tau < \text{Min}_i[-1/T_L(i)].
\label{timestep}
\end{equation}
Note that the upper bound is a
non-zero quantity for a finite system.  

On the other hand, when $i \ne j$ we have
\begin{equation}
P_{i \to j}(\tau)=-\tau H_{ij}\frac{\psi_T(j)}{\psi_T(i)} \;\;\;\; (i \ne j),
\label{pijdiff}
\end{equation}
a positive expression since $\psi_T(i)$ is chosen to be positive and 
off-diagonal terms $H_{ij}$ are negative (Eq. (\ref{tij})).

Using expressions (\ref{pii}) and (\ref{pijdiff}) 
for the transition probability random walks in configuration space can be 
generated. By averaging over configurations, statistical estimates for various 
quantities can be defined.
A first important example is the calculation of the variational energy 
associated with $|\psi_T>$ (variational Monte Carlo). The variational energy 
is defined as
\begin{equation}
E_v(\psi_T)=
\frac{<\psi_T|{\cal H}|\psi_T>}{<\psi_T|\psi_T>}.
\label{vmc}
\end{equation}
Here, it is rewritten as
\begin{equation}
E_v(\psi_T)= \lim_{K\rightarrow \infty} \frac{1}{K} \sum_{i=1}^{K} 
E_L(i) = <<E_L>>_{(P)}
\label{vmc2}
\end{equation}
where $<<\dots>>_{(P)}$ is the stochastic average over configurations $|i>$
generated using the transition probability $P$, $K$ 
being the number of configurations calculated. Equation (\ref{vmc2}) holds
because ${\psi_T(i)}^2$ is the stationary density of the stochastic process,
that is
\begin{equation}
\sum_i {\psi_T(i)}^2 P_{i \to j}(\tau) = {\psi_T(j)}^2 \;\;\;\;  \forall j.
\label{statio}
\end{equation}
This  property is directly verified by using expressions (\ref{pij}) 
and (\ref{relEL}). 

As already pointed out, the estimate of the exact energy is based on the 
stochastic calculation of ${[1-\tau({\cal H}-E_T)]}^n|\psi_T>$, 
Eq. (\ref{lim2}). Introducing between each operator in the product 
the decomposition of the 
identity over the basis set, $1=\sum_i |i><i|$, 
and making use of the definition of the transition 
probability, (\ref{pij}),  we get the following path integral representation
$$
{[1-\tau({\cal H}-E_T)]}^P \mid \psi_T \rangle = 
\sum_{i_0 \dots i_P}
{\psi_T(i_0)}^2 \prod_{k=0}^{P-1} P_{i_k \to i_{k+1}} 
$$
\begin{equation}
\prod_{k=0}^{P-1}  w_{i_k i_{k+1}}  
\frac{1}{\psi_T(i_P)} \mid i_P \rangle
\label{closrel}
\end{equation}
where the weights $w_{ij}$ are defined as follows
\begin{equation}
w_{ij} \equiv \frac{ \langle i \mid [1-\tau ({\cal H}-E_T)] \mid j \rangle}
{\langle i \mid [1-\tau ({\cal H}-E_L)] \mid j \rangle}
\label{weight1}
\end{equation}
or more explicitly,
$$
w_{ij}=1  \;\;\;\; i \ne j,
$$
\begin{equation}
w_{ii} = \frac{1-\tau(H_{ii}-E_T)}{1-\tau(H_{ii}-E_L(i))}
\;\;\;\; i = j.
\label{weight2}
\end{equation}
From (\ref{lim2}) the exact energy can be obtained as
\begin{equation}
E_0 = \lim_{P \rightarrow \infty}
\frac{ <\psi_T | {\cal H} {[1-\tau ({\cal H}-E_T)]}^P |\psi_T>}
     { <\psi_T |   {[1-\tau ({\cal H}-E_T)]}^P |\psi_T>}
\label{e0}
\end{equation}
which is rewritten here in terms of stochastic averages as
$$
E_0 = \lim_{P \rightarrow \infty}
$$
\begin{equation}
<< E_L(i_P) \prod_{k=0}^{P-1} w_{i_k i_{k+1}}>>_{(P)}/
<< \prod_{k=0}^{P-1} w_{i_k i_{k+1}}>>_{(P)}.
\label{e0qmc}
\end{equation}
 In  order to compute the averages appearing in that expression 
two strategies can be employed. 
First, formula (\ref{e0qmc}) can be directly used as it stands: 
Paths are generated using the transition probability and the local 
energy at each step is weighted by the quantity $W=\prod_k w_{i_k i_{k+1}}$. 
This approach where the number of configurations is kept fixed and the 
weights are carried out along trajectories is usually referred to as the
Pure Diffusion or Green's function Monte Carlo method. 
For extended systems such as those 
considered here, this approach is not optimal. Indeed,
it is important to sample less frequently regions 
of configuration space where the total weight is small and to accumulate 
statistics where it is large. To realize this,
a birth-death process (or branching process) 
associated with the local weight is introduced. 
In practice, it consists in adding 
to the standard stochastic move defined by the transition probability,
a new step in which the current configuration is destroyed or copied a 
number of times proportional to the local weight. Denoting 
$m_{ij}$ the number of copies (multiplicity) of the state $j$, we take
\begin{equation}
m_{ij} \equiv \text{int} (w_{ij} +\eta)
\end{equation}
where $\text{int}(x)$ denotes the integer part of $x$,
and $\eta$ a uniform random number on $(0,1)$. 
Adding a branching process can be viewed as sampling with a 
generalized transition probability ${P}^{*}_{i \to j}(\tau)$ defined as
$$
{P}^{*}_{i \to j}(\tau) \equiv P_{i \to j}(\tau) w_{ij}
$$
\begin{equation}
=\psi_T(j) \langle j \mid [1- \tau ({\cal H}-E_T)]\mid i \rangle
\frac{1}{\psi_T(i)}.
\label{pijtilde}
\end{equation}
Of course, the normalization is not constant (the population fluctuates) 
and $P^*$ is not a genuine transition probability. However, 
we can still define a stationary density for it. 
From Eq. (\ref{pijtilde}) we see that the stationary condition is 
obtained when $E_T$ is chosen to be the exact energy $E_0$, and that 
the density is $\psi_T(i) \psi_0(i)$.
Accordingly, by using a stabilized population of configurations the exact 
energy may be now obtained as
\begin{equation}
E_0 = 
{<< E_L >>}_{(P,w)}.
\label{e0dmc}
\end{equation}
Note the use of an additional subscript $w$ in the average to recall the
presence of the branching process.

At this point, we shall not expand further the method. For 
more details regarding the implementation of GFMC to lattice systems the 
interested reader is referred to Refs.\cite{carlson,gross,ceperley2,runge}.
Let us just emphasize on two important aspects.  
First, there exists a so-called zero-variance property 
for the energy: The better the trial wave function $\psi_T$ is,
the smaller the statistical fluctuations are. In the limit of an exact 
wave function for which the local energy is a constant, fluctuations entirely 
disappear (zero variance). From this important remark follows that 
in any QMC method, it is crucial to optimize as much as 
possible the trial wave function used.
Of course, in practice, a compromise 
between the complexity of the wave function and the gain in reduction 
of variance has to be found. 

Once a good trial wave function has been chosen, the only room left for 
improvement is the implementation of the dynamical process itself. 
In the algorithm presented here the only dynamical parameter which can be 
adjusted is the time-step $\tau$. 
In a configuration $|i>$ associated with a small local kinetic energy $T_L(i)$,
the system remains in this configuration a relatively large time
and a large value of $\tau$ is necessary to help the system to escape from it.
Unfortunately, because of the constraint (\ref{pii}) ($P_{i\to i }$ must be 
positive) configurations with a high local kinetic energy 
impose a small value of $\tau$.
In order to circumvent this difficulty,
we propose to integrate out exactly the time evolution of the system 
when trapped in a given configuration. This idea is developped in the 
next section.

\subsection{GFMC and Poisson processes}
Consider the probability that the system remains in a given configuration 
$i$ a number of times equal to $n$. It is given by
$$
{\cal P}_{i}(n) \equiv
P(i_1=i,\tau;\dots;i_n=i,\tau;i_{n+1}\ne i,\tau)=
$$
\begin{equation}
{[P_{i \to i}(\tau)]}^n 
[1-P_{i \to i}(\tau)]
\label{poisdisc}
\end{equation} 
${\cal P}_{i}(n)$ defines a normalized discrete Poisson distribution. 
In terms of the local kinetic energy it can be rewritten as
\begin{equation}
{\cal P}_{i}(n) = -\tau T_L(i) \exp{[ n \ln(1+\tau T_L(i)) ]}
\label{poisexpl}
\end{equation}
where the integer $n$ runs from zero to infinity. 
To describe transitions towards states $j$ different from $i$ we introduce
the following escape transition probability
\begin{equation}
{\tilde P}_{i \to j } =\frac{P_{i \to j}(\tau)}{1-P_{i \to i}(\tau)}
\;\;\;\;\; j \ne i .
\label{diftrans}
\end{equation}
Using Eqs. (\ref{pii}) and (\ref{tli}) ${\tilde P}_{i \to j }$ is rewritten in 
the most explicit form
\begin{equation}
{\tilde P}_{i \to j } =\frac{H_{ij} \psi_T(j)} 
                           {\sum_{k \ne i} H_{ik} \psi_T(k)} 
\;\;\;\;\; j \ne i .
\label{diftrans2}
\end{equation}
Note that this transition probability is positive, normalized, and 
independent of the time-step $\tau$.
Now, by using both probabilities, ${\cal P}_{i}(n)$ and 
${\tilde P}_{i \to j }$, the path integral representation of 
${G({\cal H})}^P|\psi_T>$, formula (\ref{closrel}), can be rewritten as
$$
[1-\tau ({\cal H}-E_T)]^P \mid \psi_T \rangle = \sum_{(i,n) \in {\cal C}_P}
\psi_T(i_0)^2  
$$
\begin{equation}
[\prod_{k=0}^{l-1} {\cal P}_{i_k}(n_k) {\tilde P}_{i_k \to i_{k+1}}]
{\cal P}_{i_l}(n_l)
\prod_{k=0}^{l} w_{i_k}^{n_k} 
\frac{1}{\psi_T(i_l)} \mid i_l \rangle
\label{pathdisc}
\end{equation}
where the sum is performed over the set of 
all families of states $(i_0 \dots i_l)$ 
with multiplicities $(n_0 \dots n_l)$ verifying
$\sum_{k=0}^{l-1}(n_k+1)+n_l =P$. In a given family 
successive states are different and the variable $n_k$ represents the number
of times the system remains in configuration $i_k$.
The set of all families is denoted ${\cal C}_P$ and 
an arbitrary element is written $(i,n)\equiv (i_0 \dots i_l,n_0 \dots n_l)$.
Since off-diagonal weights are equal to one, Eq. (\ref{weight2}),
a shortened notation for the diagonal weights, $w_i\equiv w_{ii}$,
has been introduced.

Now, let us remark that the time-step $\tau$ plays a role in 
the path integral formula (\ref{pathdisc}) only when the system is trapped 
into a given configuration. 
Indeed, both the escape probability 
${\tilde P}$ and the off-diagonal weight $w_{ij}$ are independent of 
$\tau$. As an important consequence the limit $\tau \rightarrow 0$
and $P \rightarrow \infty$ with $P \tau= t$ can be done exactly. In this limit 
the discrete Poisson process ${\cal P}_{i}(n)$ defined in (\ref{poisexpl}) 
converges to a continuous Poissonian distribution for the variable
$\theta=n\tau$ 
\begin{equation}
{\cal P}_i(\theta) =  \frac{1}{{\bar \theta_i}}
e^{-\theta/{\bar \theta_i} }.
\label{poiscon}
\end{equation}
In this formula ${\bar \theta_i}$ represents the average time spent 
in configuration $i$. In what follows we shall refer to it as the average 
trapping time, its expression is
\begin{equation}
{\bar \theta_i} = -1/T_L(i).
\label{trapp}
\end{equation}
The fact that ${\bar \theta_i}$ is inversely proportional to the local 
kinetic energy is explained as follows. When the kinetic energy
is small the system is almost blocked in its configuration and ${\bar \theta}$ 
is large.  In contrast, when a large kinetic energy is available, the system 
can escape easily from its current configuration and ${\bar \theta}$ 
is small. As already remarked the escape transition 
probability is independent of $\tau$ and is therefore not affected by the 
zero-time-step limit. Finally, after exponentiating the product of weights, 
the path integral can be rewritten in the form
$$
e^{-t({\cal H}-E_T)}\mid \psi_T \rangle = \sum_{i_0 \cdots i_l} 
\int_{0}^{+\infty} d\theta_0 \cdots \int_{0}^{+\infty}d\theta_l
\psi_T(i_0)^2 
$$
\begin{equation}
[\prod_{k=0}^{l-1} {\cal P}_{i_k}(\theta_k) {\tilde P}_{i_k \to i_{k+1}}]
{\cal P}_{i_l}(\theta_l)
e^{-\sum_{k=0}^l \theta_k (E_{L}(i_k)-E_T)}
\frac{1}{\psi_T(i_l)} \mid i_l \rangle
\label{pathdisc2}
\end{equation}
with the constraint that the trapping times verify
$\sum_{k=0}^l \theta_k =t$.

In order to compute ground-state properties the limit $t \rightarrow \infty$ 
must be performed, Eq. (\ref{lim2}). In this limit the constraint 
$\sum_{k=0}^l \theta_k =t$ can be relaxed and, quite remarkably, 
integrations over the Poisson distributions for the
different trapping times can be performed. 
For large enough time $t$ we obtain
$$
e^{-t({\cal H}-E_T)}|\psi_T> \sim_{l \rightarrow \infty}  
$$
\begin{equation}
\sum_{i_0 \cdots i_l}
\psi_T(i_0)^2 
\prod_{k=0}^{l-1} {\tilde P}_{i_k \to i_{k+1}} 
\prod_{k=0}^l  {\tilde w}_{i_k} \frac{-1}{T_L(i_l)}
\frac{1}{\psi_T(i_l)} \mid i_l \rangle
\label{pathdisc3}
\end{equation}
where the new integrated weights ${\tilde w}$ are found to be
\begin {equation}
{\tilde w}_i= \frac{T_L(i)}{E_T -H_{ii}}.
\end{equation}
In the same way as before the exact energy can be obtained as
\begin{equation}
E_0 = \lim_{t \rightarrow \infty}
\frac{ <\psi_T | {\cal H} e^{-t({\cal H}-E_T)} |\psi_T>}
     { <\psi_T |   e^{-t({\cal H}-E_T)} |\psi_T>}.
\label{e0ter}
\end{equation}
In terms of stochastic averages it gives
$$
E_0 = \lim_{l \rightarrow \infty}
$$
\begin{equation}
<< E_L(i_l) {\bar \theta_{i_l}} \prod_{k=0}^{l} {\tilde w}_{i_k}
>>_{({\tilde P})} /
<< {\bar \theta_{i_l}} \prod_{k=0}^{l} {\tilde w}_{i_k}>>_{({\tilde P})}.
\label{e0qmc3}
\end{equation}
where configurations are generated using the escape
transition probability ${\tilde P}$. 

As in the standard approach it is preferable to simulate the weights via
a branching process. Here also, the
reference energy $E_T$ stabilizing the population is given by the 
exact energy, $E_0$.  The new stationary density writes
\begin{equation}
\pi(i) \sim {\psi_T(i) \psi_0(i) }/{\bar \theta_i}
\end{equation}
up to an immaterial normalization constant. Finally, our estimator  for
$E_0$ is
\begin{equation}
E_0 = \frac{ << {\bar \theta_{i}} E_L(i) >>_{({\tilde P},{\tilde w})} }
{ << {\bar \theta_{i}} >>_{({\tilde P},{\tilde w})} }
\label{e0fin}
\end{equation}
where configurations are generated using ${\tilde P}$ and branched with 
${\tilde w}$.  Note that the variational energy can be recovered by removing 
the branching process (${\tilde w}=1$)
\begin{equation}
E_v(\psi_T)=
\frac{ << {\bar \theta_{i}} E_L(i) >>_{({\tilde P})} }
{ << {\bar \theta_{i}} >>_{({\tilde P})} }
\label{vmcfin}
\end{equation}

\section{Computational details}

In this section some important aspects of the practical implementation of the 
GFMC approach to the SU(N) Hubbard model are presented.

\subsection{Hard-core boson Hamiltonian}

The Hamiltonian considered here is the one-dimensional SU(N) Hubbard 
model described by (\ref{hun}).  Simulations are performed for a finite ring of 
length $L$. In one dimension the sites can be labelled in such way that 
the hopping term connects only sites represented by consecutive integers. 
As a consequence no fermion sign appears, except eventually when a fermion 
crosses the boundary ($1 \rightarrow L$ or $L \rightarrow 1$). By 
choosing either periodic or antiperiodic boundary conditions this sign can
always be absorbed and our model (\ref{hun}) becomes equivalent to a model made up
with hard-core bosons and described by
\begin{equation}
{\cal H}= -t \sum_{i=1}^{L}\sum_{a=1}^{N} c^+_{i+1 a} c_{i a} + {\rm H.c.}
+ \frac{U}{2} \sum_i ( \sum_{a} n_{i a})^2
\label{su4bos}
\end{equation}
where  $c^+_{i a}$ creates a hard-core boson of 
color $a$ on site $i$, $n_{i a}$ is 
the occupation number $n_{i a}= c^+_{i a} c_{i a}$, 
and $c^+_{L+i a} \equiv c^+_{i a}$.

\subsection{Trial wave function}

As already emphasized a most important aspect of any Monte Carlo scheme is the 
choice of a good trial wave function. To guide our choice, let us consider 
the exact solution at $U=0$.
In this case the ground-state is obtained by filling $N$ independent Fermi seas 
consisting of planes waves with momenta $k_n=2\pi n/L$ $(n=0,\pm 1,\dots)$. For a 
given type of fermion, the ground-state can be written as a 
Vandermonde determinant \cite{ogata} 
and the following expression for the ground-state is obtained
\begin{equation}
\psi_0^{U=0}(i_1,\dots i_P) =
 \prod_{l < l^\prime} \sin[\frac{\pi}{L}(i_l-i_{l^\prime})]
\end{equation}
where $i_1,\dots,i_P$ are the positions of the $P$ fermions on the chain, 
$i_k=1,\dots,L$.
In terms of occupation numbers the solution can be rewritten as
\begin{equation}
\phi(n_1,\dots,n_L) = e^{{\displaystyle \frac{^t \vec{n} {\cal A}_0
\vec{n}}{2}}}
\label{tnan}
\end{equation}
where the matrix ${\cal A}_0$ of size $(L\times L)$ is given by
\begin{equation}
{\cal A}_0(i,i^\prime)=\ln |\sin[\frac{\pi}{L}(i-i^\prime)]|.
\label{matrixa}
\end{equation}
Note that (\ref{tnan}) and (\ref{matrixa}) describe a system of particles interacting 
via a logarithmic potential (1D Log-gas). 
The exact ground-state wavefunction of the 
complete SU(N) model at $U=0$ is simply obtained by writting the 
product of the $N$ wavefunctions (\ref{tnan}) associated with each color.

When the Coulomb interaction is switched on, we have chosen to take the same
functional form as before for $\psi_T$
\begin{equation}
\psi_T(\vec{n}) \equiv e^{{\displaystyle 
\frac{^t \vec{n} A_U\vec{n}}{2}}}.
\end{equation}
Here, $A_U$ is an arbitrary matrix of size $(NL\times NL)$. Taking into account 
the translational and SU(N) symmetries, at most $L+2$ independent variational 
parameters can be defined. In all GFMC calculations presented in this paper the 
entire set of parameters has been systematically optimized. 
To do that, we have generalized the correlated sampling method of 
Umrigar {\sl et al.} \cite{umrigar} along the lines presented in the preceding 
section. To be more precise, the set of configurations used to calculate 
the quantities to be minimized (variational energy or variance of ${\cal H}$,
see Ref. \cite{umrigar}) are generated 
using the escape transition probability and 
weighted with the corresponding average trapping times. Doing this,
the effective number of configurations is increased and the 
optimization process is facilitated.
We have found that that large numbers of parameters can be easily optimized.

\subsection{$O(L)$ algorithm}

In the occupation-number representation the numerical effort 
for calculating the trial wave function $\psi_T(\vec{n})$ is of order $O(L^2)$.
To evaluate the local energy the Hamiltonian
has to be applied to the vector $|\psi_T>$. 
Since a given configuration $\mid \vec{n} \rangle$ is connected by ${\cal H}$
to about $O(L)$ states, the total computational cost per Monte Carlo step is about
$O(L^3)$.  
In fact, this cost can be reduced to $O(L)$. To do that, we introduce the 
following set of $2NL+1$ variables
\begin{equation}
(\vec{n},\vec{n}_U,n_0) 
\equiv (\vec{n},A_U{\vec{n}},\frac{^t \vec{n} A_U \vec{n}}{2}).
\label{recurrence}
\end{equation}
Using this representation, the wave function is given by $e^{n_0}$.
Configurations connected by the Hamiltonian differ from each other 
by removing a particle of a given color $a$  on a site $i$
and putting it on a neighboring site $j$.
In the occupation-number language it corresponds to add one to 
the component ${j a}$ and remove one to the component ${i a}$ of vector $\vec{n}$.
For convenience let us introduce the vector $\vec{\delta}^{(i a)}$ whose 
components 
are zero except the component ${ia}$ which is equal to one. Using
the new variables just defined we have
$$
(\vec{n},\vec{n_U},n_0) \to 
$$
$$
(\vec{n}+\vec{\delta}^{(j a)}-\vec{\delta}^{(i a)}
, \vec{n}_U + A_U\vec{\delta}^{(j a)}-A_U\vec{\delta}^{(i a)},
$$
\begin{equation}
n_0 + \frac{ ^t(\vec{\delta}^{(i a)}-\vec{\delta}^{(j a)})A_U
(\vec{\delta}^{(i a)}-\vec{\delta}^{(j a)}) }{2} 
-^t \vec{n_U}(\vec{\delta}^{(i a)}-\vec{\delta}^{(j a)}))
\label{ol}
\end{equation}

In the simulation the set of new variables is stored for each 
configuration. At each Monte Carlo step they are reactualized using (\ref{ol}). 
Finally, the numerical effort is limited to $O(L)$.

\section{Results}

Let us now present the results 
for the SU(2)-, SU(3)- and SU(4) Hubbard models. 
SU(2) results have been obtained by solving numerically the 
Lieb-Wu equations\cite{lieb}.
Other results have been obtained with the GFMC method presented in the previous 
section.  In all calculations we have set $t=1$.

\subsection{Charge gaps}

\begin{figure} \begin{center}
  \fbox{\epsfysize=5cm\epsfbox{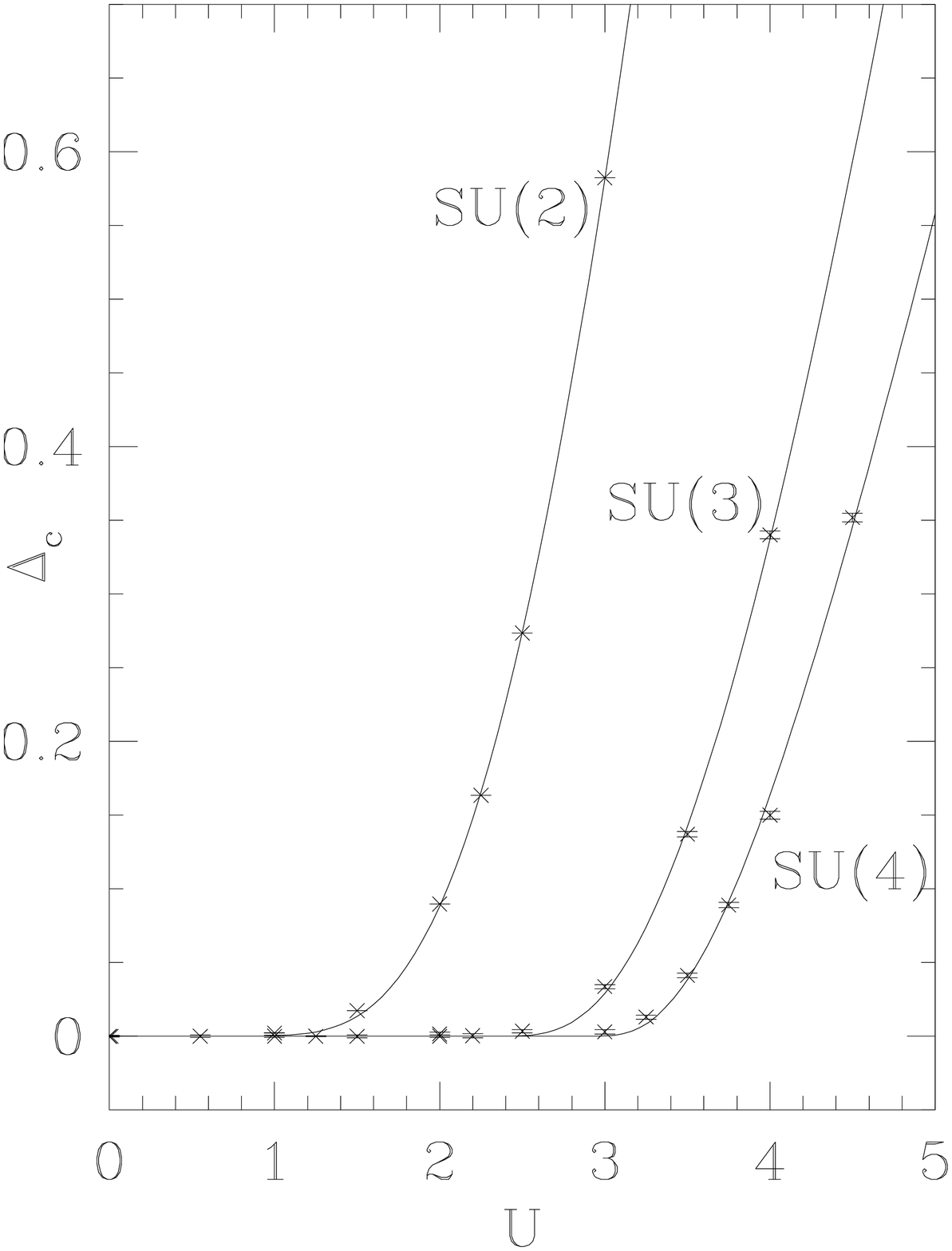}}
   \fcaption{
Charge gaps as a function of the interaction $U$ for the
SU(2), SU(3), and SU(4) Hubbard models. The values of the gaps have
been extrapolated to $L \rightarrow \infty$ (see text).}
\label{figgap}
\end{center}\end{figure}

The finite-size charge gap $\Delta_c(N_e,L)$ is defined as
$$
\Delta_c(N_e,L) \equiv 
$$
\be
E_0(N_e+1,L)+E_0(N_e-1,L)-2E_0(N_e,L)
\label{defdeltac}
\ee
where $E_0(N_e,L)$ is the total ground-state energy of a 
ring of length $L$ with $N_e$ electrons. In this expression $N_e \pm 1$ means 
that a fermion of an arbitrary color is added to or removed from the system.
Denoting $N$ the number of colors, calculations are done for a number 
of fermions of each color equal to $L/N$, and therefore for a total density
$n \equiv N_e/L$  equal to one. In order to get the 
exact charge gap the limit $L\rightarrow \infty$ must be performed. 
As usual this is done by calculating charge gaps for different sizes 
and extrapolating to infinity. Here, SU(3) and SU(4) calculations have 
been done for $L=9,12,18,27$ and  $L=8,16,24,32$, respectively. The finite-size 
gaps have been found to converge almost linearly as a function of the 
inverse of the size. Accordingly, the limit 
$L\rightarrow \infty$ of the gap has been obtained from a fit of the data 
with a linear or quadratic function of $1/L$. 
Figure \ref{figgap} presents the charge gaps 
obtained for $N=2,3,4$ as a function of the Coulomb interaction $U$.

A first important remark concerns the quality of the Monte Carlo simulations.
As it can be seen in Fig. \ref{figgap}, 
the error bars on the different gaps are quite small. 
A typical value is about 0.001. Errors are small because
total energies are calculated with a very high level of accuracy. For example,
for the SU(4)-model with $L=32$ and $U=0.5$, we get $E_0(32,32)=-52.13056(15)$
for a total number of elementary Monte Carlo steps equal to $8. 10^7$.
Clearly, the relative error of about $3. 10^{-6}$ is very small.
In the large $U$-regime where the trial wave function is not expected to be as good 
as for small $U$, we still get excellent results.
For example, for  $U=4.5$ we get $E_0(32,32)=-23.7118(13)$ ($1.6 10^8$ MC 
steps) with a relative error of about $6. 10^{-5}$. Using 
the standard GFMC method (presented in Sec. III.A) we get, for $U=0.5$,
$E_0(32,32)=-52.13050(40)$ and, for $U=4.5$, $E_0(32,32)=-23.7210(110)$ 
(in both cases the maximum time-step allowed has been chosen, see Eq. 
(\ref{timestep})).
The improvment resulting from the new approach, particularly at large $U$, 
is noticeable. Finally, using the approach of 
Trivedi and Ceperley \cite{ceperley2} (introduction of the Poisson process but 
no integration in time)  we get for $U=0.5$ $E_0(32,32)=-52.13041(22)$
and for $U=4.5$, $E_0(32,32)=-23.7121(30)$. These results illustrate the improvment  
resulting from the time integration.

Having at our disposal such accurate results we can try to
find out whether or not a gap opens for a non-zero value of $U$. 
Considering only continuous transitions, two scenarios 
are possible. A first possibility is to open a gap for any non-zero 
value of $U$. In that case we write the gap versus $U$ as follows
\be
\Delta_c= C \exp{(-G/U)}.
\label{deltac}
\end{equation}
A second scenario consists in looking for the existence of a 
KT-type transition at a finite value $U_c$ of the 
Coulomb interaction. In that case the gap is written as
\be
\Delta_c= C_{KT} \exp{ (-\frac{G_{KT}} {\sqrt{U-U_c}} ) }
\label{deltaKT}
\ee
for $U > U_c$ and zero otherwise.
The three sets of results have been fitted either using Eqs. (\ref{deltac}) or 
(\ref{deltaKT}). The fitting procedure used is a standard one, based on 
the minimization of a chi-square type function including statistical errors.
Our most important conclusion is that all sets of data can be correctly 
represented within our small statistical errors either using the gapful 
representation, Eq. (\ref{deltac}), or using a KT scenario, Eq. (\ref{deltaKT}) 
with a not too large value of $U_c$. 
For example, using Eq. (\ref{deltac}) possible representations are
($C=25.313,G=11.318$), ($C=274.634,G=26.745$), and ($C=515.649,G=32.755$),
for $N=2,3,$ and $4$, respectively. Although no clear physical
content can be given to the magnitude of coefficients, it is nevertheless satisfactory 
to verify that in the case of SU(2), the gapful (\ref{deltac}) leads to not too large
values for the coefficients. This should be contrasted with the SU(3) and SU(4) cases for 
which the parameters are important.
Within a KT scenario all data can also be very well fitted. 
In the case of SU(2) where we know for sure that no KT transition 
exists, the `critical value' issued from our fits ranges from 
$0$ to about $0.5$. For example, a possible representation is given by
($C_{KT}=541.310,G_{KT}=11.053$, and $U_c=0.384$). 
For the SU(3)-model accurate representations can be obtained with a value of
$U_c$ ranging from 0 to about 2.3  For $U_c=2.2$ (the value we shall propose later for 
the critical value) we get: ($C_{KT}=45.050, G_{KT}=6.567$, and $U_c=2.2$).
For SU(4) the interval is larger. Allowed values range from 0 to about 2.9.  
For $U_c=2.8$ (our proposed value, see below)
we get: ($C_{KT}=17.889, G_{KT}=5.144$, and $U_c=2.8$).
In contrast with the gapful representation, it should be noted that coefficients are 
now much larger for the SU(2) model than for the SU(3) and SU(4) models.

In conclusion, using accurate values of the gaps no conclusions can be 
reasonably drawn about the existence or not of a KT-type transition 
at a finite value of $U$. Numerical evidences based on other quantities are 
therefore  called for (see next sections).
From the fitting of our data the only conclusion we are allowed to draw is that 
a KT-transition is only possible within the range (0,2.3) for SU(3) and within the range 
(0,2.9) for SU(4). In addition to this, if such a transition actually occurs in both models, we 
should expect a difference for the critical values given by: $U_c[SU(4)]-U_c[SU(3)]
\sim 0.5-0.6$ (see Fig. \ref{figgap}).

\subsection{Spin gap}

The spin gap is defined as the change in ground-state energy produced 
when destroying a fermion of a given color and creating a fermion of 
a different color (in the SU(2) case it consists in flipping one spin).
Note that in this process the charge number is kept fixed.
For a finite system we have
\begin{equation}
\Delta_s(N_e,L) \equiv E_0(N_e+1,L)-E_0(N_e-1,L)
\label{defdeltas}
\end{equation}
where $N_e \pm 1$ involves an arbitrary pair of electrons of 
different colors.

\begin{figure} \begin{center}
  \fbox{\epsfysize=5cm\epsfbox{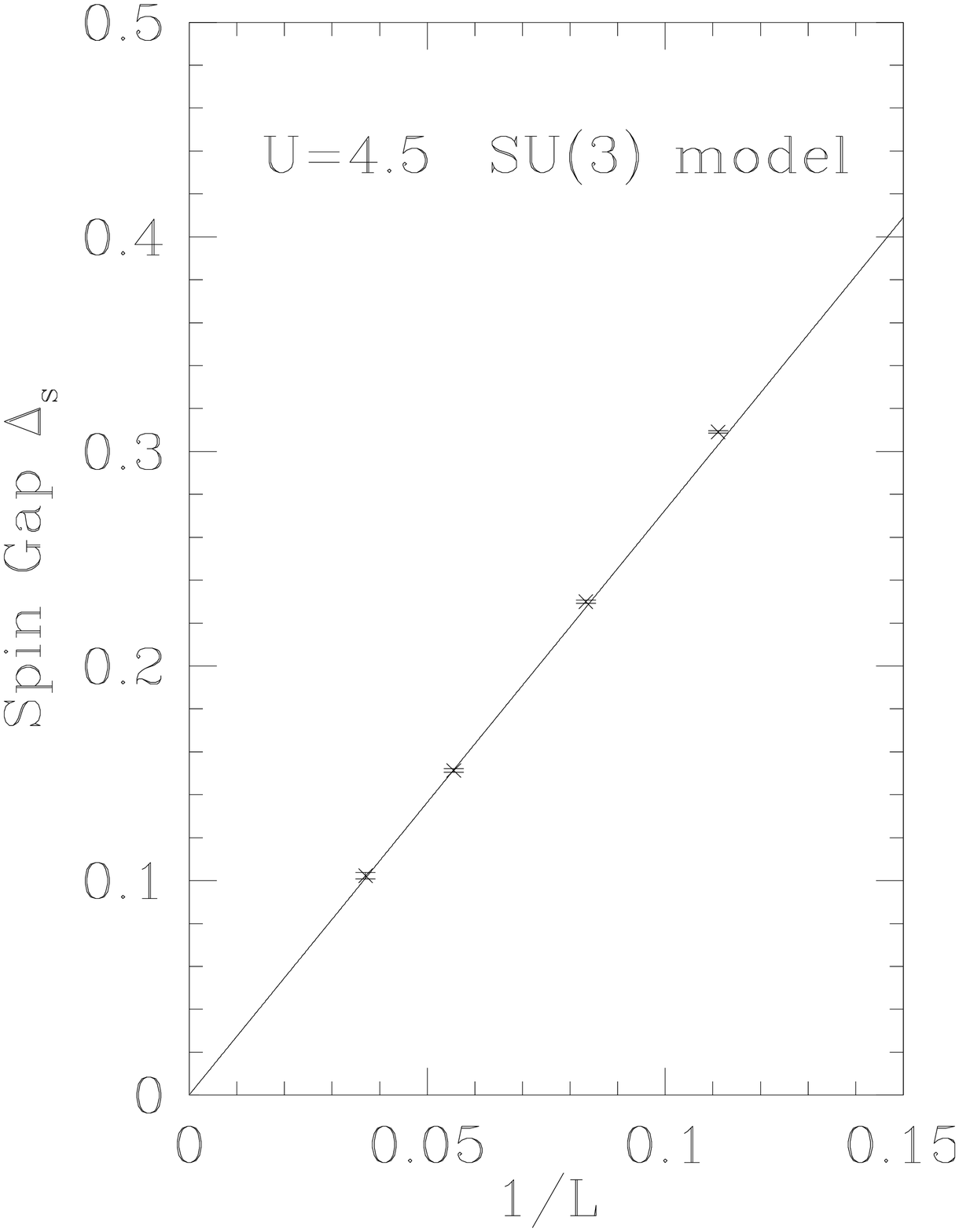}}
   \fcaption{
Spin gap as a function of $1/L$ for the SU(3) Hubbard model at $U=4.5$.
The solid line is a linear fit of the data.}
\label{figspin}
\end{center}\end{figure}

For the SU(2) case the system is known from the exact solution to be 
gapless for an arbitrary value of the interaction strength $U$. 
For a number of colors greater than 2, it is an open question.
This is an important point since the existence of a gapful regime would very 
likely indicate the existence of a coupling between spin and charge degrees of 
freedom.
In all calculations performed for $N$=3 and 4, and
for a coupling constant $U$ ranging from 
very small to very large values (up to $U=10$) no
evidence for the existence of such a gap has been found.
Thus, it can be quite safely concluded that the spin sector of SU(N) $N=2,3,4$ 
is gapless for an arbitrary interaction in full agreement with the bosonization
prediction. To illustrate this point we present in 
Fig. \ref{figspin} a typical behavior for the spin gap of SU(3) 
as a function of 
$1/L$ at the relatively large value $U=4.5$ (at least two times greater than 
the maximal value expected for $U_c$ in the charge sector). The behavior of the 
gap is essentially linear and extrapolation to the origin leads to a 
vanishing gap.

\subsection{Luttinger liquid parameters}

In this section we present calculations of the Luttinger liquid 
parameters $u_c$ and $K_c$. For that we shall make use of 
their relations with the compressibility $\kappa$ and charge stiffness 
$D_c$ of the system. For a model with $N$ colors (SU(N)) we have
the following relations
\begin{equation}
\frac{ \pi u_c} {K_c}  \kappa n^2 ={N \over 2}
\label{LTE1}
\end{equation}
and
\begin{equation}
D_c= N u_c K_c
\label{LTE2}
\end{equation}
where $n=N_e/L$ ($N_e$ total number of electrons) is the electron density.  
The compressibility ${\kappa}$ is defined as the 
second derivative of the ground-state energy $E_0$ with respect to the
density of particles
\begin{equation}
\frac{1}{\kappa}=\frac{1}{L}\frac{\partial^2 E_0}{\partial n^2}.
\label{defkappa}
\end{equation}
A convenient finite-size approximation of the compressibility is
$$
\kappa={L\over {N_e}^2}
$$
\be
\biggl({E_0(N_e+N,L)+E_0(N_e-N,L)-2E_0(N_e,L) \over
N^2}\biggr)^{-1}
\label{finitekappa}
\end{equation}
where $N_e \pm N$ in $E_0$ means that $N$ fermions -one of each color-
are added to or removed from the system.

The charge stiffness is given by
\begin{equation}
D_c= {\pi \over L} \left .
\frac{\partial^2 E_0}{\partial \varphi^2}
\right|_{\varphi=0}
\label{defdrho}
\end{equation}
\noindent where $\varphi$ is a charge twist in the system. This charge twist
is imposed by introducing the following twisted boundary conditions 
\be
c^+_{i+L a}= e^{i\varphi} c^+_{i a},
\ee
for an arbitrary site $i$ and color $a$.

By calculating with GFMC total ground-state energies for different numbers of 
electrons, formula (\ref{finitekappa}) allows a direct calculation of the 
compressibility.
In contrast, the GFMC calculation of the charge stiffness is more tricky
due to the presence of a complex hopping term 
at the boundary. To circumvent this difficulty we resort to the second-order 
perturbation-theory expression of the charge stiffness. We have
\begin{equation}
D_c= 
{\pi \over L} \biggl( <-T> - 2 \sum_{k\ne 0} \frac{{|<k|J|0>|}^2}
                                                   {E_k-E_0}
 \biggr)
\label{drhopert}
\end{equation}
where $T = -t \sum ( c^+_{i+1 a} c_{i a} + H.c.)$ is the 
kinetic-energy operator, $J= -i t \sum 
( c^+_{i+1 a} c_{i a} - H.c.)$ is the paramagnetic current operator,
$<\cdots>$ denoting the expectation value in the ground-state, all quantities 
being evaluated at $\varphi=0$. To evaluate the kinetic term we make use of
the Hellman-Feynman theorem: $<T> = E_0 - U
\frac{\partial E_0}{\partial U}$. In practice, the following 
finite-difference expression is used
\begin{equation}
<T> = E_0 - U \biggl( \frac{ E_0(U+\delta U) -E(U-\delta U)} {2 \delta U}
 \biggr)
\label{Tfinite}
\end{equation}
with $\delta U$ small enough to make higher-order contributions  
negligible. 

The second-order part of formula (\ref{drhopert}) can 
be re-interpreted back as the second-derivative of the total ground-state 
energy of a new Hamiltonian consisting of the original Hamiltonian plus 
a perturbation associated with the flux operator $J$.
This leads to the following relation
\begin{equation}
\sum_{k\ne 0} \frac{{|<k|J|0>|}^2} {E_0-E_k}
= \frac{1}{2} \frac{\partial^2 \tilde{E_0}(\lambda)}{\partial \lambda^2}
\label{E2deriv}
\end{equation}
where $\tilde{E_0}$ is the ground-state energy of the new Hamiltonian defined
by
\begin{equation}
\tilde{H} =  -(t+\lambda) \sum_{ia} ( c^+_{i+1 a} c_{i a} )
             -(t-\lambda) \sum_{ia} ( c^+_{i-1 a} c_{i a} ) +V(U)
\label{Htilde}
\end{equation}
and $V(U)$ is the potential part of the problem. 
Using formulas (\ref{E2deriv}) and (\ref{Htilde}) 
the charge stiffness can now be obtained from a series 
of GFMC ground-state calculations of total energies of {\sl real} Hamiltonians
(more precisely, $E_0$, $E_0(\delta U)$, and $E_0(-\delta U)$ for 
${\cal H}$, and $\tilde{E_0}(\lambda)$ for $\tilde{H}$, Eq. (\ref{Htilde})).
It should be emphasized that the new Hamiltonian $\tilde{H}$ is real but not 
symmetric: Left-moving and right-moving electrons do not have the same velocity. Of 
course, such a property is easily implemented within a QMC framework.

\begin{figure} \begin{center}
  \fbox{\epsfysize=5cm\epsfbox{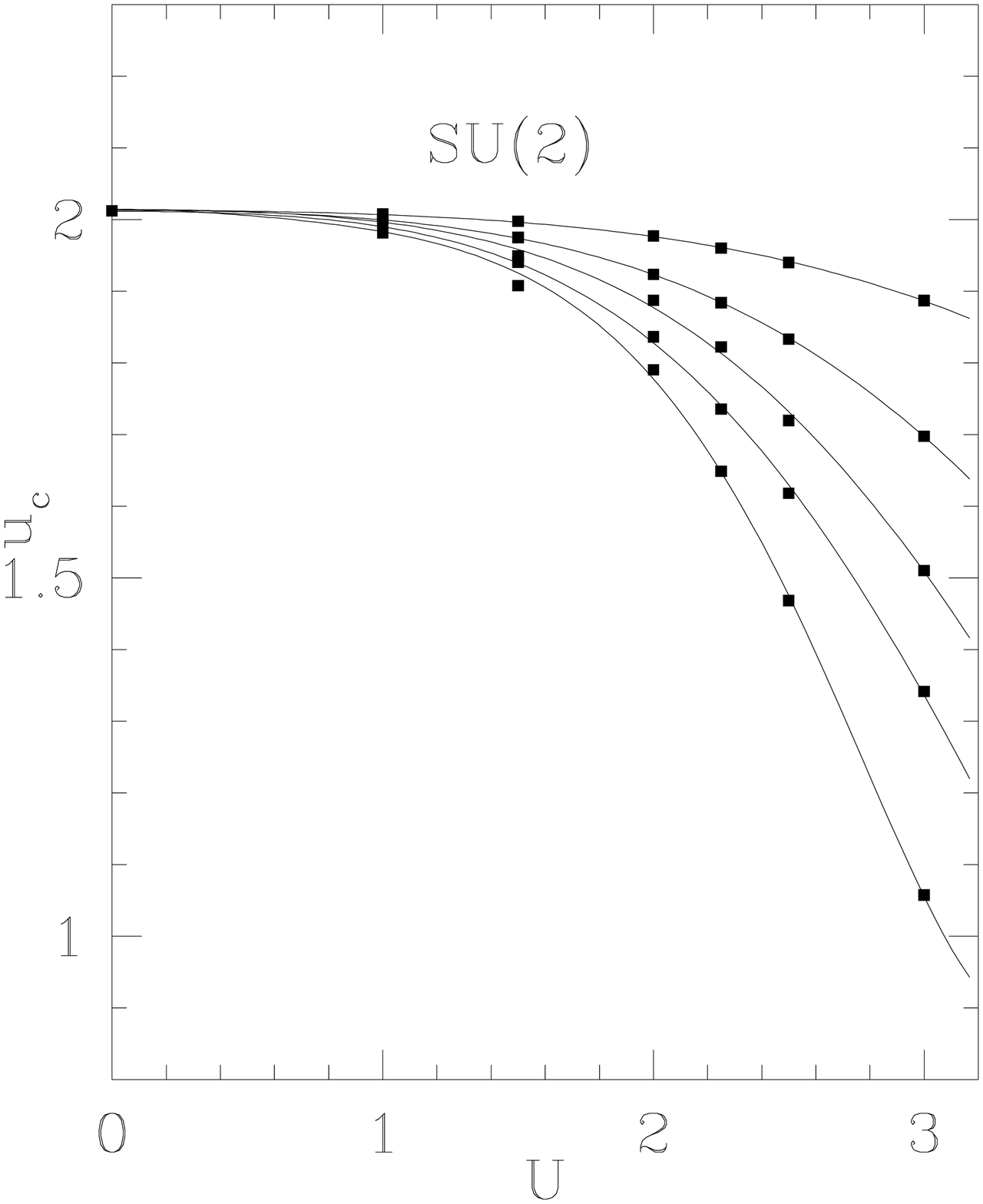}}
   \fcaption{
$u_{c}$ as a function of $U$ for the SU(2) Hubbard model.}
\label{figUcSU2}
\end{center}\end{figure}

\begin{figure} \begin{center}
  \fbox{\epsfysize=5cm\epsfbox{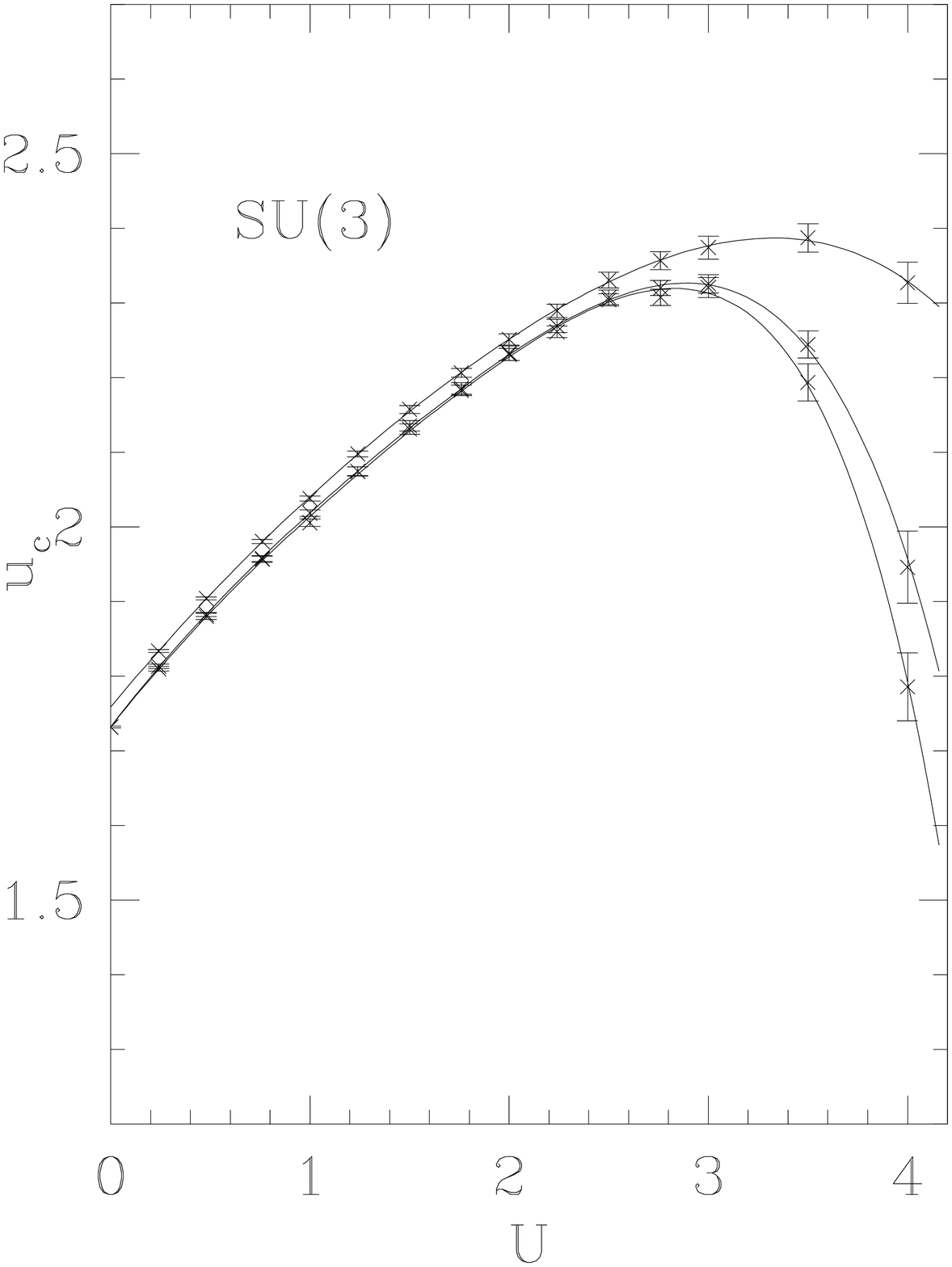}}
   \fcaption{
$u_{c}$ as a function of $U$ for the SU(3) Hubbard model.}
\label{figUcSU3}
\end{center}\end{figure}

\begin{figure} \begin{center}
  \fbox{\epsfysize=5cm\epsfbox{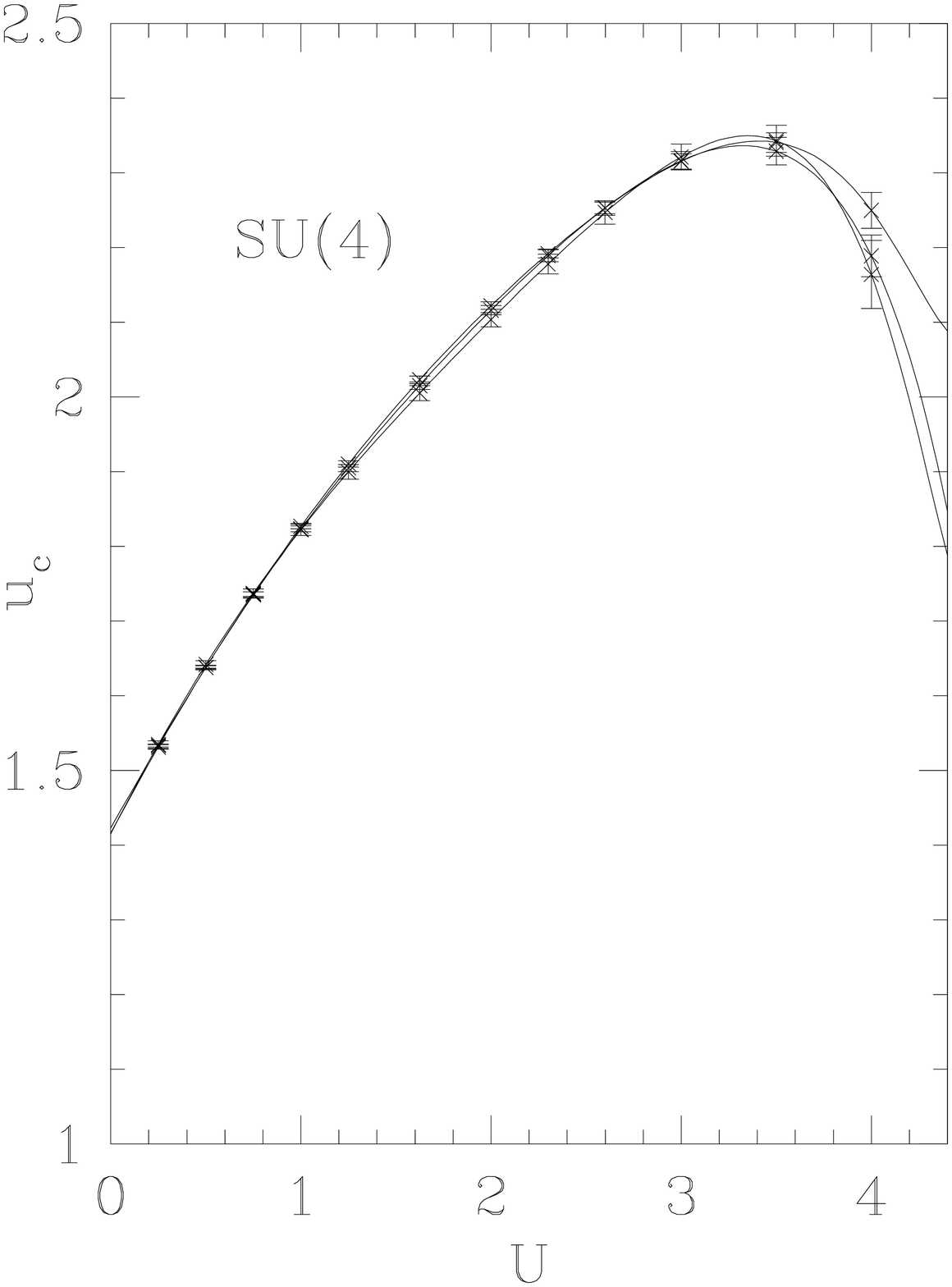}}
   \fcaption{
$u_{c}$ as a function of $U$ for the SU(4) Hubbard model.}
\label{figUcSU4}
\end{center}\end{figure}

\begin{figure} \begin{center}
  \fbox{\epsfysize=5cm\epsfbox{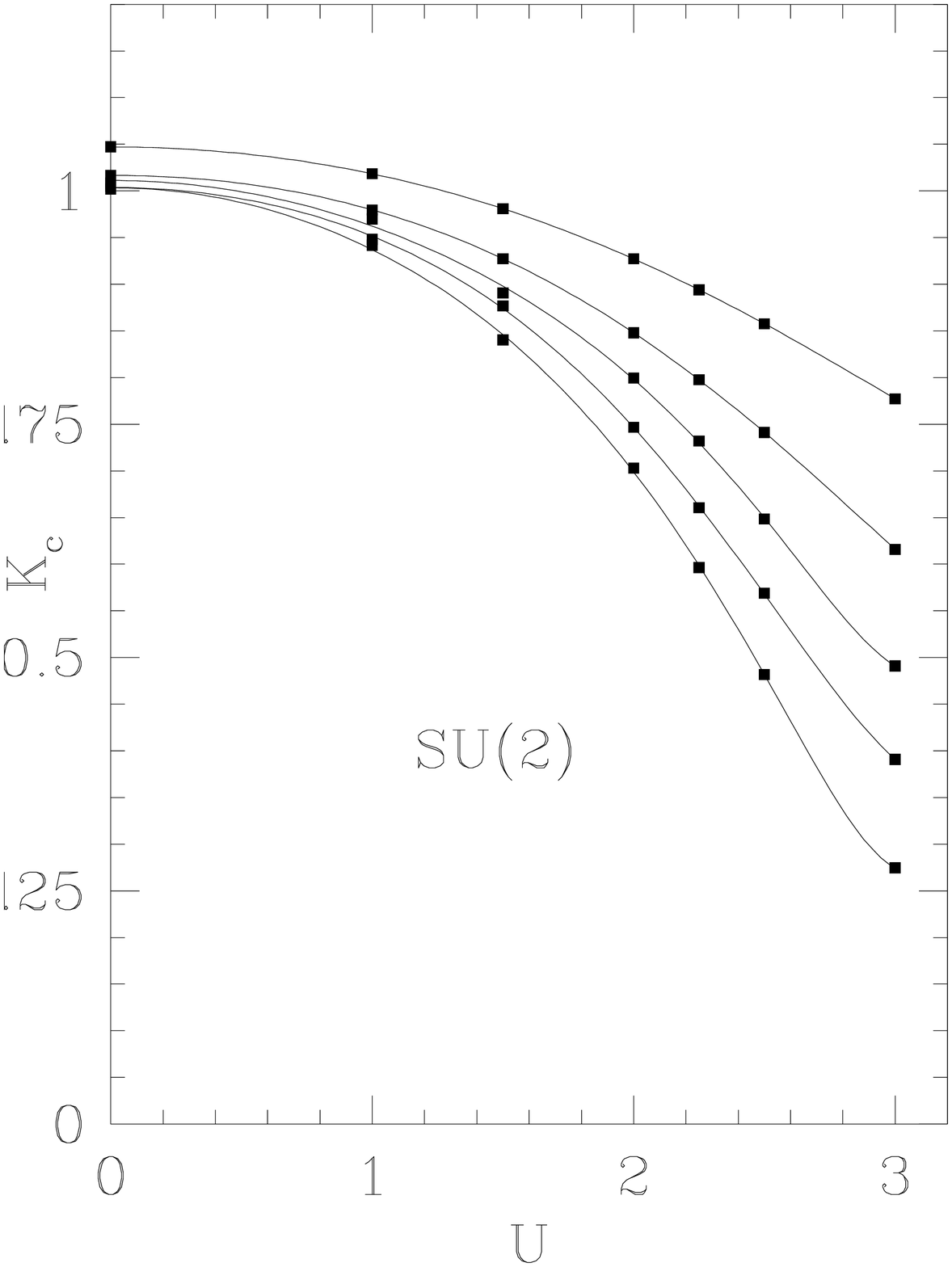}}
   \fcaption{
$K_{c}$ as a function of $U$ for the SU(2) Hubbard model.}
\label{figKcSU2}
\end{center}\end{figure}

\begin{figure} \begin{center}
  \fbox{\epsfysize=5cm\epsfbox{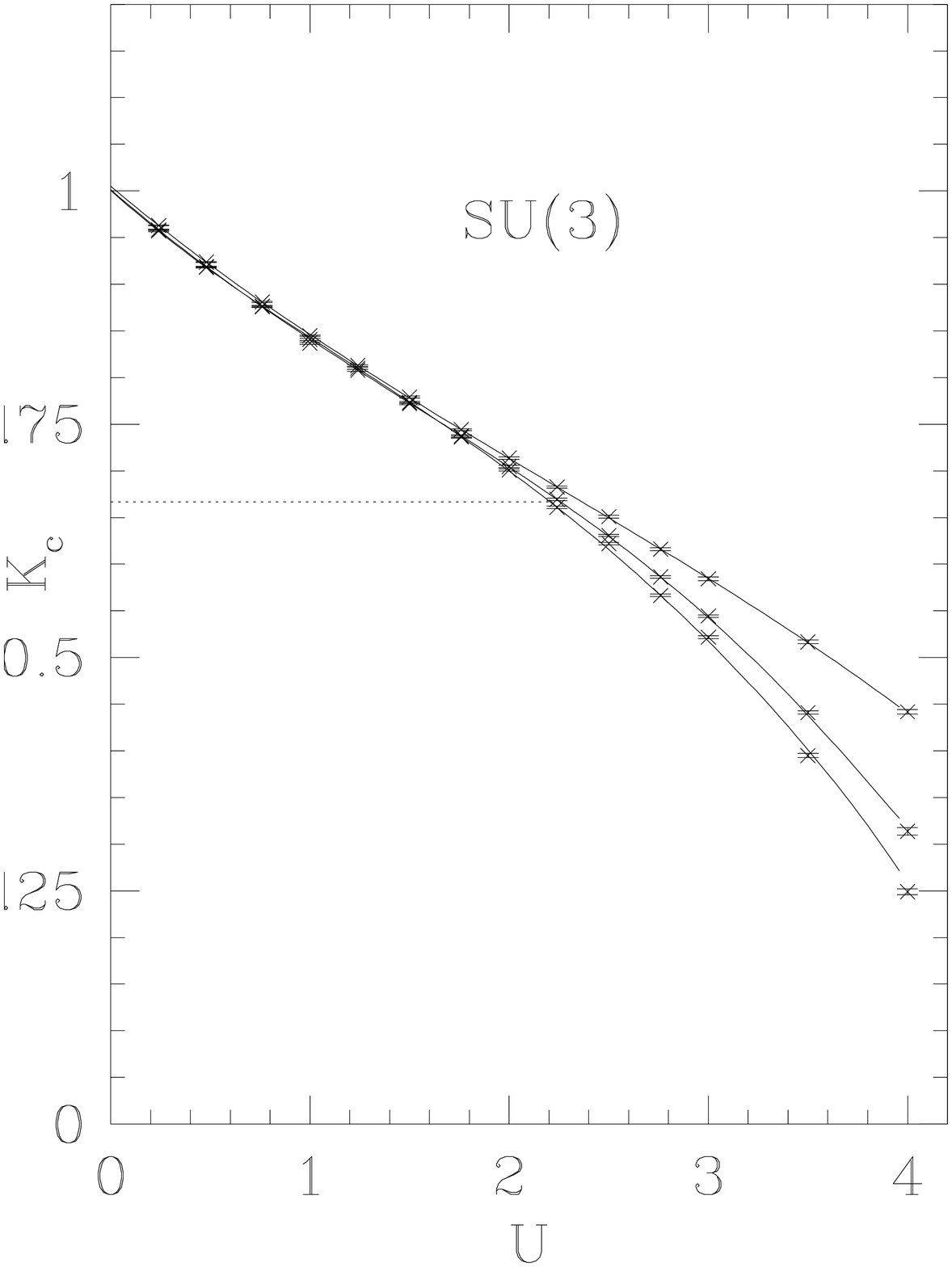}}
   \fcaption{
$K_{c}$ as a function of $U$ for the SU(3) Hubbard model.}
\label{figKcSU3}
\end{center}\end{figure}

\begin{figure} \begin{center}
  \fbox{\epsfysize=5cm\epsfbox{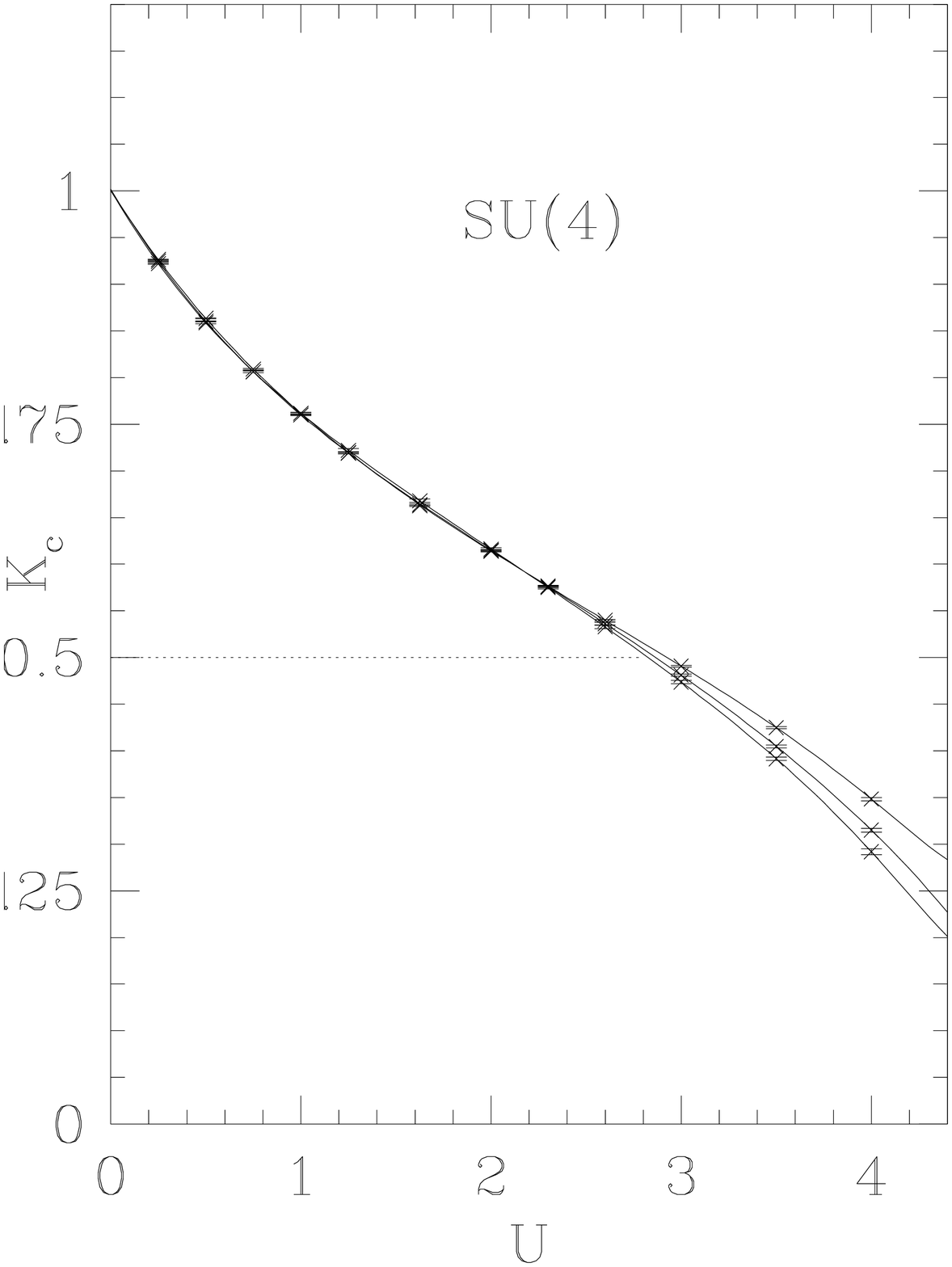}}
   \fcaption{
$K_{c}$ as a function of $U$ for the SU(4) Hubbard model.}
\label{figKcSU4}
\end{center}\end{figure}

Figures 
[\ref{figUcSU2},\ref{figUcSU3},\ref{figUcSU4},\ref{figKcSU2},\ref{figKcSU3},\ref{figKcSU4}] 
present the Luttinger parameters $u_c$ and $K_c$ for the SU(2), SU(3), and 
SU(4) Hubbard models as a function of the interaction $U$ and for different sizes $L$.
For the SU(2) model, parameters have been obtained by computing ground-state 
energies issued from the standard Lieb-Wu equations (computation of the 
compressibility, formula (\ref{finitekappa})) and
from their generalization to the case of twisted boundary conditions as presented by 
Shastry and Sutherland \cite{shastry90} (computation of the charge stiffness,
formula (\ref{defdrho})). For the SU(3) and SU(4) models we have followed the general route just 
presented above. 

A first striking result emerging from the figures is the strong qualitative differences 
between the general behavior of Luttinger parameters of the SU(2) model 
on the one hand, and 
of the SU(3) and SU(4) models, on the other hand. Let us first have a look at the charge 
velocity $u_c$.

In the SU(2) case the charge velocity has been calculated for various values of $U$ and for the 
sizes $L=6,10,14,18,$ and $22$. Results are presented in 
Fig. \ref{figUcSU2}. The
upper curve corresponds to $L=6$, the lower one to the maximum size, $L=22$. In between, the 
curves are ordered according to the magnitude of $L$. 
For a given size $L$, the charge velocity is found to decrease as a function of 
$U$. For a given $U$, $u_c$ also decreases as a function of the size $L$.
Such a behavior is quite typical of a gapped system in which collective charge 
excitations are damped away. In the limit of large 
sizes, the charge velocity is expected to vanish for a non-zero value of the interaction.

The charge velocities of the SU(3) model, Fig. \ref{figUcSU3}, and of the SU(4) model,
Fig. \ref{figUcSU4}, display a very similar behavior which is dramatically 
different from the one 
observed for SU(2). Starting from their free value at $U=0$ ($u_c=\sqrt{3}$ and 
$u_c=\sqrt{2}$ for SU(3) and 
SU(4), respectively), they increase as a function of $U$ with a finite slope at 
the origin. After some critical value of $U$ both 
velocities go down quite rapidly. In the first part 
of the curves (small and intermediate values of $U$) the charge velocity is found 
to converge quite rapidly as a function of the size. All curves presented 
cannot be distinguished within statistical errors. Although the calculations 
presented here are limited to systems with a maximum size of 
$L=27$ (SU(3)) or $L=32$ (SU(4)) some preliminary calculations at larger 
sizes strongly suggest that the values plotted are indeed converged.  
Such results strongly support the existence of 
a gapless phase for the SU(3) and SU(4) models. At larger values of $U$ the situation is rather 
different. The charge velocities decrease quite rapidly both as a function of $U$ and as a function 
of $L$. This behavior indicates the existence of a gapped phase. In order to be more
quantitative let us have a look at the value of the slope at the origin. The theoretical 
prediction can be obtained from Eqs. (\ref{luttparaper}). For SU(3) 
calculations have been done for the sizes L=9,18, and 32.
The slope at the origin is found to be
0.32(1),0.32(1), and 0.33(2) for L=9,18, and 27, respectively.
Theses results are in perfect agreement with the theoretical prediction of
$\frac{1}{\pi} \simeq 0.318$. For the SU(4) model
calculations have been done for the sizes L=16,24, and 32.
The slope at the origin is found to be
0.46(1),0.47(1), and 0.45(2) for L=16,24, and 32, respectively.
Here also, the results are in perfect agreement with the theoretical prediction of
$\frac{3}{2\pi} \simeq 0.477$. 
Let us now consider our results for $K_c$. 

Here also, there exists a common behavior
for the cases SU(3) and SU(4), and a different one for SU(2). In the latter 
case, Fig. \ref{figKcSU2}, $K_c$ decreases either as a function of $U$ or 
as a function of the size. The slope 
at the origin, $U=0$, is essentially zero and $K_c$ is expected to vanish at large sizes, 
except, of course, in the free case. Once again, this behavior is typical of a 
gapped system.
In the two other cases, the situation is rather different. In the same way as for the 
charge velocity, two regimes can be distinguished, see Figs. \ref{figKcSU3} 
and \ref{figKcSU4}. At small and intermediate $U$, the 
values of $K_c$ are found to be very well converged within statistical errors 
as a function of the size $L$. The curve is smooth with a finite slope at the origin. 
In the second regime corresponding to larger values of $U$ the curves $K_c$ versus $U$ 
go down as a function of the size. Clearly, this latter regime corresponds to a 
gapped phase. Having 
nearly  exact values of $K_c$ up to some critical value $U_c$ for SU(3) and 
SU(4), the next logical step consists in comparing these values to the predictions of 
bosonization. A first important prediction was the opening of a gap in the charge 
sector for a value of $K_c$ equal to $2/N$, Eq. (\ref{Kcrt}). In Fig. \ref{figKcSU3}
corresponding to the SU(3) case, a dashed line has been drawn at the value $K_c=2/3$. 
The intersection of this line with the curves of $K_c$ appears at 
about $U_c \sim 2.2$.  A most remarkable result
is that this value of $U$ is both consistent with the critical value 
extracted from the calculation of the charge gaps, Fig. \ref{figgap}, 
but also with the fact that it lies in the domain of $U$ where the values of 
$K_c$ begin not to converge as a function of the size 
(a fact usually interpreted as resulting from 
the existence of a finite correlation length).
A very similar situation is obtained in the SU(4) case. Using the same 
type of arguments, $U_c$ is 
found to be around 2.8. When studying charge gaps we had observed a 
difference of $U_c$, Fig. \ref{figgap}, between SU(3) and SU(4) of between 0.5 and
0.6. This is in very good agreement with
what is found here from independent data on $K_c$. A second prediction 
which can be tested is the estimate of the value of $U_c$ itself. 
Formula (\ref{ucrt}) gives
$$
U_c = \frac{\pi}{2} \; \frac{N^2 - 4}{N-1} \; \sin\frac{\pi}{N}.
$$
For $N=3$ and $N=4$ one gets $U_c=3.40$ and $U_c=4.44$, respectively. As already 
pointed out, these estimates must be considered with caution.
However, it should give the correct trend as a function of $N$. 
Here, if we look at the ratio $U_c[SU(4)]/U_c[SU(3)]$ we get about 1.31 from 
the theoretical estimate and about 1.27 from our data. 
The agreement is excellent.
Another point which can be checked is the value of the slope at the origin. For the 
SU(3) case, it is found to be -0.18(1),-0.19(1), and -0.19(2) for L=9,18, and 27, respectively.
Theses results are in very good agreement with the theoretical prediction of
$-\frac{1}{\sqrt{3}\pi} \simeq -0.183$ given by Eq. (\ref{luttparaper}).
For SU(4) we find a slope of -0.31(1),-0.33(1), and -0.32(2) for L=16,24, and 32, respectively. 
Theses results are also in total agreement with the theoretical prediction of 
$-\frac{3}{2\sqrt{2} \pi} \simeq -0.337$. 

Finally, it can be very useful for interested readers to give
some compact and accurate representations of the Luttinger parameters $K_c$ and $u_c$ as 
a function of $U$. For both parameters a minimal 
representation we may think of (see section II.B)
is 
\bea
K_c &=& \frac{1}{ \sqrt{1 + k_1 U + k_2 U^2 }} \nonumber \\
u_c  &=& v_F \sqrt{1 + u_1 U + u_2 U^2 }.
\label{luttparafit}
\eea
For SU(3) we obtain 
$$
k_1=0.33452, k_2=0.08789
$$
$$
u_1=0.37929, u_2=-0.025509.
$$
Note that these values are not 
too far from the bare values corresponding to Eqs. (\ref{luttpara}),
$k_1^0=u_1^0=\frac{2}{\pi v_F} \simeq 0.36755$, $k_2^0=u_2^0=0$.

For SU(4) we obtain 
$$
k_1=0.62065, k_2=0.12298
$$
$$
u_1=0.71486, u_2=-0.052705
$$
to compare to the bare values given by
$k_1^0=u_1^0=\frac{3}{\pi v_F}=0.675237$, $k_2^0=u_2^0=0$.

As already discussed we have found no evidences for the opening of 
a spin gap in the case of the SU(3) and SU(4) models. In other words, 
the system remains critical with respect to the spin degrees of freedom  
for any value of the interaction. For these models the slope at
the origin is predicted to be equal to
$-\frac{1}{2\pi} \simeq -0.159$ (Eq. (\ref{velo})). Once again,
this value has been recovered using our numerical data. To compute the 
spin velocity we have used the formula expressing the spin gap as a 
function of the size for a critical system \cite{korepin}
\begin{equation}
u_s = \frac{\Delta_s(N_e,L)} {2\pi L}.
\label{defus}
\end{equation}
For SU(3) and SU(4) we get for the slope -0.18(2) and -0.18(3), respectively,
in very good agreement with the theoretical prediction.

\begin{figure} \begin{center}
  \fbox{\epsfysize=5cm\epsfbox{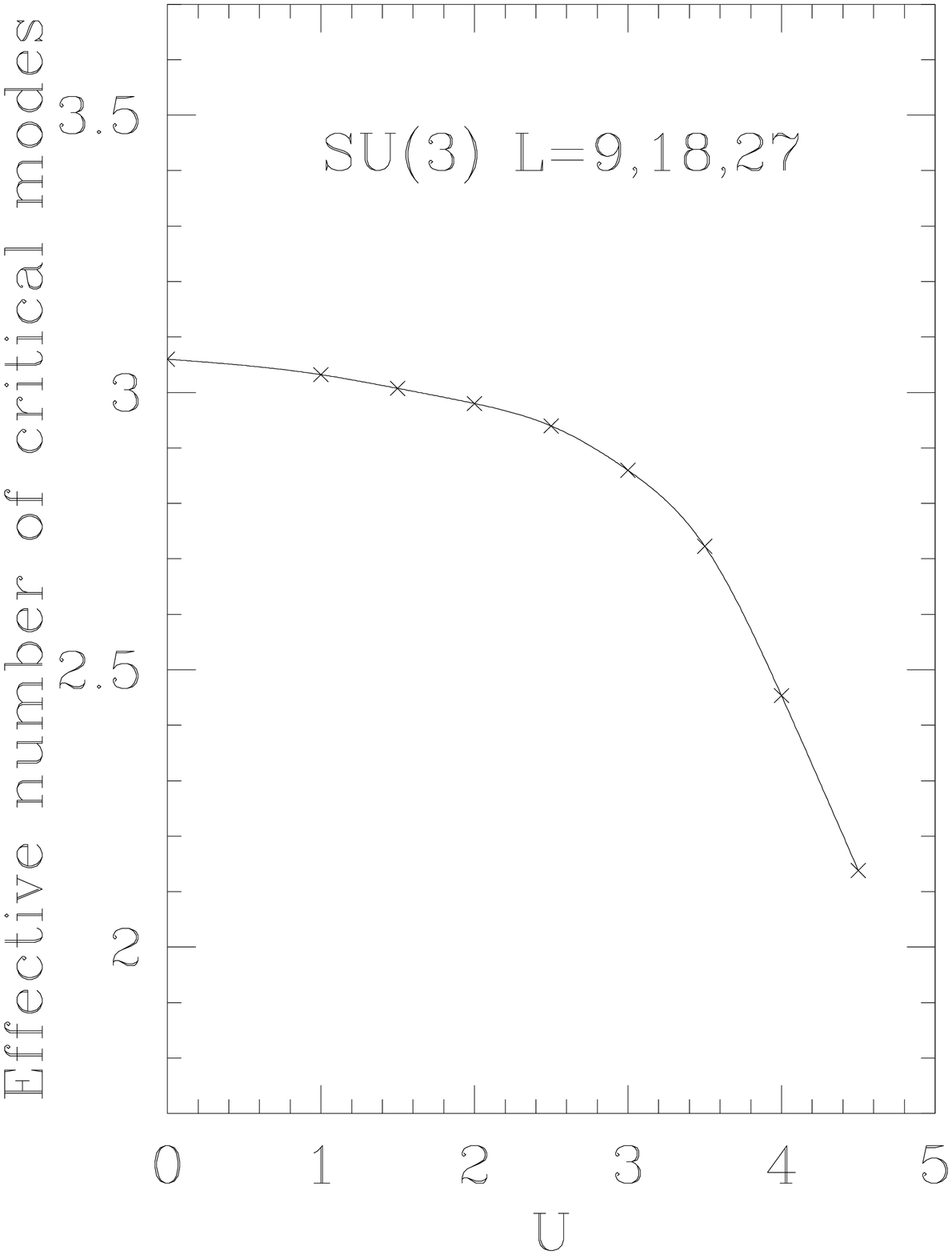}}
   \fcaption{Effective number of critical modes as a function of $U$ 
for the SU(3) Hubbard model.}
\label{figceffSU3}
\end{center}\end{figure}

A final piece of information which can be extracted from our data is related to the 
way the total ground-state energy converges to its asymptotic value. 
To be more precise, it is known that the ground-state energy per site,
$e_0(L)$, of a Luttinger liquid is expected to behave as follows \cite{korepin}
\be
e_0(L) \simeq e_0(+\infty) - \frac{\pi}{6 L^2} \sum_i u_i 
\label{central}
\ee
where $\sum_i u_i$ denotes the total velocity associated with all critical excitations. 
In the free case,  $N$ degres of freedom are critical, and 
the total velocity is equal to $N v_F$. When the interaction is turned on, it is 
possible to follow the evolution of the total velocity as a function of $U$.
This has been done for the SU(3) model. Taking our data for the sizes L=9, 18, 
and 27 the ground-state energy has been fitted with a form adapted to
Eq. (\ref{central}), $e_0=a-b/L^2$. From this 
fit an effective number of critical modes can be defined as follows
$$
N_{eff}=\frac{6b}{\pi v_F}.
$$
The result is presented in Fig (\ref{figceffSU3}).

Although the transition is not as sharp as for the Luttinger parameters, the 
loss of one critical mode (passing from 3 to 2) is clearly seen when $U$ varies 
from zero to infinity. A similar curve may be obtained for the SU(4) case.

\section{Concluding Remarks}

In this work, we have studied the SU(N) generalization of 
the one-dimensional Hubbard model for repulsive interaction at  
half-filling. Using a combination of bosonization 
and QMC results, we have clearly shown that 
the SU(N) Hubbard model for $N >2$ behaves very differently from
the SU(2) case. Strong numerical and theoretical 
evidences have been given in favor of a Mott transition, 
between a metallic and an insulating phase,
occuring for a finite value of the Coulomb repulsion $U_c >0$
for $N>2$.

The picture emerging from the bosonization approach 
consists in a spin-charge separation at low energy.
The spin degrees of freedom are critical for arbitrary $U$
and described by the SU(N)$_1$ WZNW model with a central charge $c=N-1$
($N-1$ gapless bosonic modes). The effective theory associated 
with the charge degrees of freedom corresponds to a sine-Gordon 
model at $\beta^2 = 4\pi N K_c(U)$. 
For a small value of the Coulomb interaction $U$, the 
interaction  is irrelevant. The charge 
sector is then critical and described by a massless bosonic field. 
In this weak coupling phase, the system is metallic 
with anomalous power law behaviors in the physical quantities typical
of a Luttinger liquid.
For a finite value of the interaction $U_c$ such that 
$K_c(U_c)=2/N$, 
a KT phase transition to an insulating phase 
is expected in the bosonization approach. 
In this strong-coupling phase, the charge bosonic field  
becomes locked and the infinite discrete $Z_{\infty}$ symmetry
related to the periodicity of the potential of the
sine-Gordon model is spontaneously broken.
The only degrees of freedom that remain critical in this strong
coupling phase are the $N-1$ spin modes and 
after integrating out the massive charge degrees of freedom, 
the low-energy theory of the model corresponds to the SU(N) Heisenberg
antiferromagnet.

Very accurate numerical simulations
based on a generalization of the GFMC method and 
fully optimized trial wave functions have been performed
to obtain the spin and charge gaps,
and the Luttinger liquid parameters as 
a function of the Coulomb interaction
for the SU(2), SU(3), and SU(4) Hubbard models.
A metal-insulator phase transition at 
a finite value $U_c$ is clearly seen 
for SU(3) ($U_c \sim 2.2$) and SU(4) ($U_c \sim 2.8$) 
in contrast with the standard SU(2) case.
In addition all the results obtained for $N=3$ and $N=4$
are fully consistent with the theoretical framework 
drawn in section II.  
This provides an accurate test of the bosonization approach
to the SU(N) Hubbard model for small and large values of $U$. 
It is therefore natural to expect that
the physical picture emerging from the two
cases studied here can be extended to {\it arbitrary} values of $N$. Thus one may 
conclude that the occurrence at a finite value of the interaction
of a Mott transition of the KT type is  
{\it generic} in the SU(N) Hubbard model for $N>2$ at 
half-filling. In addition, it should be emphasized that the
calculations of the Luttinger parameters $K_c$ and $u_c$  
presented in Sec. II.B are of very good quality (in particular they are converged 
as a function of the size)
and thus provide an accurate  
characterization of the low-energy properties of the metallic phase of 
the SU(3) and SU(4) Hubbard models. 

Let us now compare our results with the exact solution of the
integrable model based on the SU(N) 
generalization of the Lieb-Wu Bethe ansatz equations \cite{schlottman}.
As discussed in the introduction, an exact solution of an SU(N)
generalization of the Hubbard model is available. Although the underlying lattice
Hamiltonian of the model is not known,  it involves very likely long-range
interactions that dynamically exclude three-electron configurations.
The question that naturally arises is whether the physics described by
the latter model is similar, when  N$>2$, to that of the
lattice SU(N) Hubbard model that we have studied in this paper.
At half-filling, the SU(N) integrable model undergoes
a $first$-order phase-transition,
as one varies $U$, from a metallic to an insulating phase\cite{frahm}.
This is in disagreement with the KT transition predicted by our analysis. In the metallic
phase the integrable model is a Luttinger liquid for every $N$
\cite{frahm,kawakami1} with the same physical properties as those obtained by
the bosonization approach for the SU(N) Hubbard model.
However, the charge stiffness $K_c$ obtained from the Bethe ansatz equations
varies between $1/N$ and $1$ as $U$ decreases from $U_c$ to $0$\cite{frahm,kawakami1}.
The value at the transition ($K_c=1/N$) is thus two times larger than the value
obtained for the SU(N) Hubbard model. This clearly confirms that the integrable
model differs from the lattice SU(N)
Hubbard model in the charge sector. As already pointed out, this difference should
result from the presence of non-local interactions in
the lattice model associated with the integrable SU(N) model.

Regarding perspectives, it is clearly of interest to 
further explore the phase diagram of the SU(N) Hubbard 
model: case of an attractive interaction, 
dependence on the filling, etc... 
For an attractive interaction at half-filling, bosonization predicts that a 
phase transition should also occur as $|U|$ varies.
For incommensurate fillings, it is easy to see, within the 
bosonization framework, that the system is a Luttinger liquid for arbitrary 
$N$ and positive $U$ where the leading asymptotics of the electronic Green's function
and spin-spin correlation coincide with those computed in the 
metallic phase.
The situation is less clear for commensurate fillings 
$k_F= \pi n/(Na_0)$ ($N/n$ being an integer).
In the bosonization approach,  
a gap opens in the charge sector for $K_c = 2n^2/N$.
The existence of a Mott transition for commensurate fillings
clearly requires the full knowledge of $K_c(U,n)$
of the lattice model.
Some preliminary calculations show that there is a very special commensurate filling,
$n=N/2$, where  no Mott transition exists and 
for which the charge and spin degrees of freedom are 
massive for $N>2$ and arbitrary $U$\cite{assaraf}.   

Let us end by noting a very interesting connection bewteen 
the metal-insulator transition predicted  in the 
SU(N) Hubbard model and the existence of plateaux in 
magnetization curves of spin ladders under a strong
magnetic field\cite{oshikawa,totsuka,cabra}. 
Using the Jordan-Wigner transformation, one can indeed
interpret the SU(N) Hubbard model as a N-leg S=1/2 XY spin ladder
in an uniform magnetic field along the z-axis and 
coupled in a symmetric way by Ising interaction. 
The relation between the Fermi momenta 
and the magnetization $<M>$ (normalized
such that the saturation value is $\pm 1$) 
is: $k_F = \pi (1-<M>)/(2a_0)$.
The Mott transition found in this work for the SU(N) Hubbard model 
at half-filling corresponds to the appearance of plateaux 
at $<M> = (N-2)/N$ in the magnetization curves of the previous N-leg XY spin 
ladder. Moreover, the existence of a Mott transition for 
the SU(N) Hubbard model at commensurate filling will give additional 
plateaux located at  $<M> = (N-2n)/N$ in the magnetization 
curves of the corresponding spin ladder. 

{\bf Acknowledgments}:
We would like to thank 
A. O. Gogolin and A. A. Nersesyan for valuable discussions
and for a related collaboration.
This work was supported by the ``Centre National de la Recherche Scientifique''
(C.N.R.S.)

\appendix

\section*{}

In this Appendix, we give some details of computations to   
establish the separation of spin and charge (\ref{spinchargesepar}) 
at the Hamiltonian level in 
the continuum limit of the SU(N) Hubbard model and fix
the expressions of $u_{c,s}$ and $G_{c,s}$ given 
by Eqs. (\ref{velo}, \ref{coupl}). 

\subsection{Sugawara form of the free Hamiltonian}

To begin with, we shall recall
some basic things on the SU(N) non-Abelian bosonization
(for a review see Refs.\cite{affleck,tsvelik,difrancesco}).
As seen in section IIA, 
the chiral SU(N) spin current ${\cal J}^A_{R,L}$ can 
be expressed in terms of $N$ left-right moving fermions $\psi_{aR,L}$: 
\be
{\cal J}^A_{R(L)} =
:\psi^{\dagger}_{aR(L)}{\cal T}^A_{ab}\psi_{bR(L)}:.
\label{currferapp}
\ee
The left (respectively right)-moving fermions are holomorphic (respectively 
anti-holomorphic) fields of the complex coordinate ($z = \tau + \ri x$,
$\tau$ being the imaginary time): $\psi_{aL}(z),
\psi_{aR}(\bar z)$. These fields are defined by the following OPEs:
\bea 
\psi_{aL}^{\dagger}\left(z\right) \psi_{bL}\left(\omega\right) \sim 
\frac{\delta_{ab}}{2\pi\left(z-\omega\right)} \nonumber \\
+ :\psi_{aL}^{\dagger}\psi_{bL}: \left(\omega\right)  
+\left(z-\omega\right) 
:\partial \psi_{aL}^{\dagger} \psi_{bL}: \left(\omega\right) +.. \nonumber \\
\psi_{aR}^{\dagger}\left(\bar z\right) \psi_{bR}\left(\bar \omega\right) \sim 
\frac{\delta_{ab}}{2\pi\left(\bar z- \bar \omega\right)} \nonumber \\
+ :\psi_{aR}^{\dagger}\psi_{bR}: \left(\bar \omega\right)
+ \left(\bar z-\bar \omega\right) 
:\bar \partial \psi_{aR}^{\dagger} \psi_{bR}: 
\left(\bar \omega \right) +.. \nonumber \\
\label{defope}
\eea
with $\partial = \partial_{\omega}, \bar \partial = \partial_{\bar \omega}$
and there are no singularities in the OPE when one does the fusion of two 
operators belonging to different sectors.

Let us now consider the OPE between two left SU(N) spin currents
for instance: 
\bea
{\cal J}^A_{L}\left(z\right) {\cal J}^B_{L}\left(\omega\right)
= :\psi^{\dagger}_{aL}{\cal T}^A_{ab}\psi_{bL}:\left(z\right)
:\psi^{\dagger}_{dL}{\cal T}^B_{de}\psi_{eL}:\left(\omega\right) \nonumber \\
= {\cal T}^A_{ab} {\cal T}^B_{de} \psi^{\dagger}_{aL}\left(z\right) 
\psi_{eL}\left(\omega\right) \psi_{bL}\left(z\right)
\psi^{\dagger}_{dL}\left(\omega\right).
\eea
Using the OPEs (\ref{defope}), the commutation relation (\ref{commulie}), 
and the normalization 
of the generators of the SU(N) Lie algebra, one obtains:
\bea
{\cal J}^A_{L}\left(z\right) {\cal J}^B_{L}\left(\omega\right) \sim 
\frac{\delta^{AB}}{8\pi^2 \left(z - \omega\right)^2} \nonumber \\
+ \frac{\ri f^{ABC}}{2\pi \left(z - \omega\right)}
{\cal J}^C_{L}\left(\omega\right).
\label{commspinl}
\eea
In the same way, we find for the right spin current:
\bea
{\cal J}^A_{R}\left(\bar z\right) {\cal J}^B_{R}\left(\bar \omega\right)\sim 
\frac{\delta^{AB}}{8\pi^2 \left(\bar z - \bar \omega\right)^2} \nonumber \\
+ \frac{\ri f^{ABC}}{2\pi \left(\bar z - \bar \omega\right)}
{\cal J}^C_{R}\left(\bar \omega\right).
\label{commspinr}
\eea
Evaluating these OPE at equal time, one recovers the OPE (\ref{OPEspin})
showing that ${\cal J}^A_{R,L}$ are SU(N)$_1$ spin current. 
With the same procedure,  one can compute the OPE between the 
charge current ${\cal J}^{0}_{R,L}$ using its definition (\ref{chargedens}) in 
terms of the underlying fermions:
\bea
{\cal J}^0_{L}\left(z\right) {\cal J}^0_{L}\left(\omega\right) &\sim& 
\frac{N}{4\pi^2 \left(z - \omega\right)^2} \nonumber \\
{\cal J}^0_{R}\left(\bar z\right) {\cal J}^0_{R}\left(\bar \omega\right)
&\sim& 
\frac{N}{4\pi^2 \left(\bar z - \bar \omega\right)^2} 
\label{commcharge}
\eea
so that the charge current ${\cal J}^{0}_{R,L}$ belongs to the  
U(1)$_N$ KM algebra.

The next step is to obtain the Sugawara 
form (\ref{sugawaraspin}, \ref{sugawaracharge}) of the 
free part of the Hamiltonian (${\cal H}_0$).
Let us consider, for instance, the left sector of the theory since
we shall obtain the same result for the right part with
the substitution: $L\rightarrow R$, 
$(z,w) \rightarrow (\bar z,\bar \omega)$ 
and $\partial \rightarrow \bar \partial$ .
We need now the following OPE for the spin sector:
\bea
{\cal J}^A_{L}\left(z\right) {\cal J}^A_{L}\left(\omega\right)
= :\psi^{\dagger}_{aL}{\cal T}^A_{ab}\psi_{bL}:\left(z\right)
:\psi^{\dagger}_{dL}{\cal T}^A_{de}\psi_{eL}:\left(\omega\right) \nonumber \\
= \frac{1}{2}\left(\delta_{ae} \delta_{bd} - \frac{1}{N}
\delta_{ab} \delta_{de}\right)
\psi^{\dagger}_{aL}\left(z\right)
\psi_{eL}\left(\omega\right) \psi_{bL}\left(z\right)
\psi^{\dagger}_{dL}\left(\omega\right)
\eea
where we have used the relation (\ref{geniden}).
Using (\ref{defope}) and keeping also the first regular terms in the fusion,
we get: 
\bea
&{\cal J}^A_{L}&\left(z\right) {\cal J}^A_{L}\left(\omega\right)
\sim  \frac{1}{8\pi^2 \left(z - \omega\right)^2} \nonumber \\
&+& \frac{N+1}{2N} :\psi^{\dagger}_{aL} \psi_{aL}
\psi_{bL} \psi_{bL}^{\dagger}:\left(\omega\right) \nonumber \\
&-& \frac{N^2-1}{2\pi N} :\psi^{\dagger}_{aL} 
\partial \psi_{aL}:\left(\omega\right).
\eea
Therefore, one obtains: 
\bea
:{\cal J}^A_{L}{\cal J}^A_{L}: =
 \frac{N+1}{2N} :\psi^{\dagger}_{aL} \psi_{aL}
\psi_{bL} \psi_{bL}^{\dagger}: \nonumber \\
- \frac{N^2-1}{2\pi N} :\psi^{\dagger}_{aL} 
\partial \psi_{aL}:.
\label{sugaspinl}
\eea
In the same way, we obtain for the left charge current:
\bea
:{\cal J}^0_{L}{\cal J}^0_{L}: =
- :\psi^{\dagger}_{aL} \psi_{aL}
\psi_{bL} \psi_{bL}^{\dagger}:
- \frac{1}{\pi} :\psi^{\dagger}_{aL} 
\partial \psi_{aL}:.
\label{sugachargel}
\eea
One can eliminate the four fermions terms by considering
the following combination:
\bea
\frac{\pi}{N} :{\cal J}^0_{L}{\cal J}^0_{L}: + \frac{2\pi}{N+1} 
:{\cal J}^A_{L}{\cal J}^A_{L}: = - :\psi^{\dagger}_{aL} \partial \psi_{aL}:.
\label{sugaHl}
\eea
Since one has $\partial \psi_{aL} = -i\partial_x \psi_{aL}$
within our convention, 
the identity (\ref{sugaHl}), the so-called Sugawara form,
states that the free Hamiltonian of N relativistic left-moving fermions 
can be written only as a function of left current-current terms.
In the right part, we have also a similar identity: 
\bea
\frac{\pi}{N} :{\cal J}^0_{R}{\cal J}^0_{R}: + \frac{2\pi}{N+1}
:{\cal J}^A_{R}{\cal J}^A_{R}: = -\ri 
:\psi^{\dagger}_{aR} \partial_x \psi_{aR}:.
\label{sugaHr}
\eea
Collecting all terms, we finaly obtain the 
Sugawara form of the free Hamiltonian ${\cal H}_0$ (\ref{h0}):
\bea
-\ri \left(:\psi^{\dagger}_{aR} \partial_x \psi_{aR}: - 
:\psi^{\dagger}_{aL} \partial_x \psi_{aL}:\right) = 
\frac{\pi}{N} \left(:{\cal J}^0_{R}{\cal J}^0_{R} \right.
\nonumber \\
\left.
+ {\cal J}^0_{L}{\cal J}^0_{L}:\right)
+ \frac{2\pi}{N+1}\left(:{\cal J}^A_{R}{\cal J}^A_{R} + 
{\cal J}^A_{L}{\cal J}^A_{L}:\right).
\label{sugafin}
\eea

\subsection{Sugawara form of the SU(N) Hubbard Hamiltonian}

We shall now investigate the effect of the Hubbard interaction 
in the continuum limit to fix the expressions 
(\ref{velo}) and (\ref{coupl})
of the 
velocities ($u_{c,s}$) and the coupling constants ($G_{c,s}$).
Using the continuum description of the SU(N) spin 
density (\ref{spincont}), the interacting part (\ref{intspin}) 
is given by dropping all oscillatory contributions:
\bea 
{\cal V}_0 = -\frac{Ua_0 N}{N+1} \left(:{\cal J}^A: :{\cal J}^A:
+:{\cal N}^A: :{\cal N}^{A \dagger}: + \right. \nonumber \\
\left. 
:{\cal N}^{A \dagger}: :{\cal N}^{A}: \right).
\label{vinterapp}
\eea
The OPE between the $2k_F$ parts of the spin density can
be computed using (\ref{2kfcurrfer}) and (\ref{defope}) as in the previous
subsection. We find up to constant terms:
\bea 
&:{\cal N}^A: &\left(z,\bar z\right) 
:{\cal N}^{A \dagger}:\left(\omega,\bar \omega\right) 
+
:{\cal N}^{A \dagger}:\left(z,\bar z\right) 
:{\cal N}^{A}:\left(\omega,\bar \omega\right) \nonumber \\ 
&\sim & 
-\frac{N^2-1}{2\pi N} \frac{z-\omega}{\bar z - \bar \omega}
:\psi_{aL}^{\dagger} \partial \psi_{aL}:\left(\omega\right) \nonumber \\
&-& \frac{N^2-1}{2\pi N} \frac{\bar z- \bar \omega}{z - \omega}
:\psi_{aR}^{\dagger} \bar \partial \psi_{aR}:\left(\bar \omega\right)
\nonumber \\
&-& :\psi_{aL}^{\dagger} \psi_{aL} 
\psi_{bR}^{\dagger} \psi_{bR}:\left(\omega, \bar \omega\right)
\nonumber \\
&+& \frac{1}{N} :\psi_{aL}^{\dagger} \psi_{bL} 
\psi_{bR}^{\dagger} \psi_{aR}:\left(\omega, \bar \omega\right).
\eea 
Using Eqs. (\ref{sugaspinl}, \ref{sugachargel}) and similar 
equations 
in the right sector
together with the definition of the charge current (\ref{chargedens}), 
we end with:
\bea 
&{\cal J}^A& {\cal J}^A
+{\cal N}^A {\cal N}^{A \dagger} + 
{\cal N}^{A\dagger}{\cal N}^{A} \sim 
-\frac{N^2-1}{2N^2}\left(:{\cal J}^0_R {\cal J}^0_R 
+ {\cal J}^0_L {\cal J}^0_L:\right)
\nonumber \\
&+& \frac{1}{N}\left(:{\cal J}^A_R {\cal J}^A_R 
+ {\cal J}^A_L {\cal J}^A_L:\right)
+ 2\frac{N+1}{N} {\cal J}^A_R {\cal J}^A_L 
\nonumber \\
&-& \frac{N^2-1}{N^2} {\cal J}^0_R {\cal J}^0_L.
\eea
As a consequence, the continuum limit of 
the SU(N) Hubbard model at half filling exhibits the 
spin-charge separation:
\be
{\cal H}= {\cal H}_c + {\cal H}_s
\label{spinchargeseparapp}
\ee
with
\be
{\cal H}_c = \frac{\pi v_c}{N} \left(:{\cal J}^0_{R}{\cal J}^0_{R}:
+ :{\cal J}^0_{L} {\cal J}^0_{L}: \right)
+ G_c \; {\cal J}^0_{R}{\cal J}^0_{L}
\label{hchargeapp}
\ee
and
\be
{\cal H}_s = \frac{2\pi v_s}{N+1} \;
\left(:{\cal J}_{R}^A {\cal J}_{R}^A: + :{\cal J}_{L}^A {\cal J}_{L}^A:
\right) + G_s \; {\cal J}_{R}^A{\cal J}_{L}^A.
\label{hspinapp}
\ee
The renormalized velocities are given by:
\bea
v_s &=& v_F - \frac{Ua_0}{2\pi} \nonumber \\
v_c &=& v_F + (N-1)\frac{Ua_0}{2\pi}
\label{veloapp}
\eea
whereas 
the current-current 
couplings in the charge and the spin sectors write:
\bea
G_c &=& \frac{N-1}{N} \; U a_0\nonumber \\
G_s &=& - 2 U a_0 .
\label{couplapp}
\eea

\end{document}